\documentclass[11pt,a4paper]{article}
\pdfoutput=1
\usepackage{jcappub}

\usepackage[english]{babel}
\usepackage{xspace} 
\usepackage{url}

\usepackage{cosmodefs.JCAP}

\usepackage{amsmath}
\usepackage{amssymb}
\usepackage{ifthen,xcolor}
\usepackage{multirow}
\usepackage{lscape}
\usepackage[size=normalsize]{caption}
\usepackage{subcaption}
\usepackage{caption}
\usepackage{graphicx}

\begin{document}

\date{\today}
\title{Probing Two-Field Open Inflation by Resonant Signals in Correlation Functions}
\author{Thorsten Battefeld,}
\emailAdd{tbattefe(AT)astro.physik.uni-goettingen.de}
\author{Jens C.~Niemeyer,}
\emailAdd{niemeyer(AT)astro.physik.uni-goettingen.de}
\author{Dimitar Vlaykov}
\emailAdd{vlaykov(AT)astro.physik.uni-goettingen.de}
\affiliation{Institute for Astrophysics,
University of Goettingen,
Friedrich Hund Platz 1,
D-37077 Goettingen, Germany}

\abstract{We derive oscillatory signals in correlation functions in two-field open inflation by means of the in-in formalism; such signatures are caused by resonances between oscillations in the tunnelling field and fluctuations in the inflaton during the curvature dominated, intermediate and subsequent inflationary regime.  While amplitudes are model-dependent, we find distinct oscillations in the power and bi-spectrum that can act as a direct probe of the curvature dominated phase and thus, indirectly, strengthen the claim of the string landscape if they were observed. We comment on the prospects of detecting these tell-tale signs in current experiments, which is challenging, but not impossible.

At the technical level, we pay special attention to the applicability conditions for truncation fluctuations to the light (inflaton) field and derive upper limits on the oscillation amplitude of the heavy field. A violation of these bounds requires a multi-field analysis at the perturbed level.} 

\maketitle

\section{Introduction}

The existence of a landscape in string theory with a plethora of metastable vacua suggests the presence of eternal inflation, yet our universe must have undergone around sixty e-folds of subsequent slow-roll inflation  to solve the problems of the big bang and lead to the observed fluctuations in the cosmic-microwave background radiation. These two regimes of inflation are connected via quantum tunnelling. If the transition can be described by a Coleman-De Luccia instanton \cite{Coleman:1980aw}, the universe in the nucleated bubble appears as an infinite, open universe. Spatial curvature is subsequently reduced below current observational bounds by slow-roll inflation. Hence, this model of open inflation \cite{Bucher:1994gb,Linde:1995xm,Yamamoto:1995sw,Linde:1995rv,GarciaBellido:1997te} seems to be a generic occurrence in string theory, providing (in principle) a theory of the initial conditions for the slow-roll phase at the background and the perturbed level.  Two observational tell-tale signs of open inflation are well known: 
spatial curvature, see e.g.~\cite{Guth:2012ww} for a recent analysis, and super-curvature fluctuations of varying origin \cite{Tanaka:1993ez,Tanaka:1994qa,Sasaki:1994yt,Yamamoto:1996qq} (see also \cite{Yamamoto:1994te,Hamazaki:1995dy,Yamamoto:1995tk,GarciaBellido:1995rv,Garriga:1996pg}) that affect correlation functions on the largest scales \cite{Yamamoto:1995sw}.

In this paper, we investigate a third indicator that is generically present if the tunnelling field $\sigma$ is not identical to the inflaton as in \cite{Linde:1995xm,Linde:1995rv}, leading to quasi-open inflation \cite{GarciaBellido:1997te}: since $\sigma$ is usually displaced from the inflationary valley after tunnelling, as in the concrete case study of \cite{Sugimura:2011tk}, or because it got excited subsequently, it exhibits oscillations during the curvature dominated phase that result in additional oscillatory signatures in correlation functions, whose frequency in k-space, $\omega(k)$, acts as a direct probe of the background scale factor and thus the curvature dominated regime.

The use of signals induced by oscillating heavy fields to identify the scale factor's evolution was first advocated by Chen \cite{Chen:2010xka,Chen:2011zf,Chen:2011tu,Chen:2012ja}, particularly to differentiate slow-roll inflation from proposed  alternatives, such as bouncing cosmologies. We extend  Chen's results to open inflation, for which ome of the approximations made in \cite{Chen:2011zf,Chen:2011tu} are not applicable.

After  introducing a simple two-field open inflation model \cite{Sugimura:2011tk} as the background of interest in Sec.~\ref{sec:twofielopen}, we turn to perturbations: following the philosophy of \cite{  Chen:2011zf,Chen:2011tu}, we truncate perturbations to fluctuations in the inflaton, and treat oscillations in $\sigma$ as a small correction to the background. We pay special attention to the validity range of such a truncation, resulting in strict upper bounds on the allowed oscillation amplitude of $\sigma$, see Sec.~\ref{sec:truncating}. These bounds are often not stated explicitly in the literature (see however \cite{Burgess:2012dz,Shiu:2011qw}). While some case studies in \cite{Chen:2011zf,Chen:2012ja} contain amplitudes in excess of our bounds, the qualitative point of these studies remains valid. If larger amplitudes are to be considered, the full multi-field setup needs to be solved at the perturbed level. The effective-field theory proposed in \cite{Achucarro:2010jv,Achucarro:2010da,Cespedes:2012hu,Achucarro:2012sm,Achucarro:2012yr,Achucarro:2012fd} has a different validity range, which does not cover the regime we are interested in (see Sec.~\ref{sec:truncating} for a brief discussion).

Given such a truncation, we use the in-in formalism 
 \cite{Schwinger:1960qe}, briefly summarized in Sec.~\ref{sec:overvieinin} (see \cite{Baumann:2009ds} and \cite{Adshead:2009cb} for pedagogical reviews), to compute correlation functions in Sec.~\ref{sec:powerandbiinin}: resonances between the mode functions and oscillations in background quantities caused by $\sigma(t)$ lead to oscillatory signals. Since these resonances concern only perturbations deep inside the horizon, we are justified to use the Bunch-Davies state as an approximation to the (excited) state after tunnelling, Sec.~\ref{sec:bd vs non-bd}.

Based on the stationary phase approximation, we derive analytic expressions for the power-spectrum and the bi-spectrum in three smoothly connected regimes, curvature domination (Sec.~\ref{sec:curv dom}), inflation (Sec.~\ref{sec:infl dom}) and an intermediate regime (Sec.~\ref{sec:com contr}). We recover Chen's results during inflation and provide concrete case studies for signals generated during the prior two regimes, Sec.~\ref{sec:casestudy} and Sec.~\ref{sec:anothercasestudy}.

We find that while an initial displacement of $\sigma$ from the inflationary valley directly after tunnelling leads to distinct oscillations in correlations functions, their amplitude is below observational sensitivity in both the power- and bi-spectrum in current experiments (within our computational strategy). However, a subsequent displacement of $\sigma$ in the curvature dominated regime, which can take place without spoiling the validity of our approximations (Sec.~\ref{sec:excitation}), can lead to signals at the percent level in the power-spectrum, if conditions are favourable ($\sigma$ is heavy and it is displaced shortly before the intermediate regime). Such signals are within reach of observations and should be sought after. The amplitude of the bi-spectrum remains  small.
We discuss our results in Sec.~\ref{sec:conclusion}.

\section{Two-field Open Inflation \label{sec:twofielopen}}
In open inflation \cite{Bucher:1994gb,Linde:1995xm,Yamamoto:1995sw,Linde:1995rv,Yamamoto:1995sw}, the slow-roll regime is preceded by a false vacuum driven phase (eternal inflation), which is terminated by a Coleman-De Luccia tunnelling event \cite{Coleman:1980aw} (see \cite{Colemanbook} for a textbook treatment). From within the nucleated bubble, the Universe can be described as an infinite, open FRW universe \footnote{In the original treatment by Coleman and De Luccia (CdL), the bubble expands almost at the speed of light  after it has nucleated; the inner edge of its wall traces out hyperbolic surfaces in the surrounding Minkowski space (in the bubble interior) which are asymptotic to the light cone at the origin of the bubble. These surfaces of constant (negative) spatial curvature are precisely where the field is constant. They subsequently correspond to surfaces of constant CMB temperature, which is our definition of constant cosmic time. Thus, the whole of the observable (infinite) universe today fits 
in the space-time interior of the finite bubble. In the two-field model that we consider the universe inside the bubble is actually finite \cite{GarciaBellido:1997te}, which is sometimes referred to as quasi-open inflation.}. Around $60$ e-folds of slow-roll inflation are needed to reduce curvature below current bounds, $\Omega_k =0.0027^ {+0.0038}_{-0.0038}$ \cite{Hinshaw:2012fq} (degeneracies with regard to the type of dark energy are lifted by current data \cite{Okouma:2012dy}).

False vacua are abundant on the landscape of string theory \cite{Susskind:2003kw}, rendering open inflation an appealing candidate on this higher dimensional field space with around $500$ moduli fields. After the tunnelling event, the fields do not find themselves generically in another local minimum; if the target region happens to be flat enough, a subsequent phase of slow-roll inflation can result. 

From a phenomenological point of view, both large- and small-field models are possible; the latter ones may be more appealing, since they avoid the $\eta$ problem and appear to be natural on ``random'' landscapes, for which slow-roll inflation usually takes place in the vicinity of an inflection point, see e.g.~\cite{Agarwal:2011wm,McAllister:2012am,Frazer:2011tg,Battefeld:2012qx,Yang:2012jf,BlancoPillado:2012cb};  a short duration of inflation is more likely than longer ones \cite{Battefeld:2012qx,Yang:2012jf} (the probability often scales as $N^{-3}$ \cite{Freivogel:2005vv,Agarwal:2011wm}, where $N$ is the number of e-folds). If this phase lasts less than around sixty e-folds, curvature disrupts subsequent structure formation, preventing the genesis of galaxies and thus observers \cite{Freivogel:2005vv}. Hence, based on mild anthropic reasoning, we expect curvature to be close to observational bounds \cite{Hinshaw:2012fq,Okouma:2012dy}, leading to an indirect observational test of the landscape idea \cite{Guth:2012ww}\footnote{This line of reasoning is hampered by the measure problem \cite{Olum:2012bn,Schiffrin:2012zf} (see \cite{Freivogel:2011eg} for a review or proposed measures). The earliest attempt at making quantitative predictions for $\Omega_k$ in open inflation can be found in \cite{Vilenkin:1996ar}, and for quasi-open inflation \cite{GarciaBellido:1997te} in \cite{Garriga:1998px}.}.

In addition to searching for spatial curvature directly, one may observe other tell-tale signs of open inflation: slight deviations from a nearly scale invariant power-spectrum on the largest scales, which arise due to the presence of super-curvature perturbations \cite{Tanaka:1993ez,Tanaka:1994qa,Sasaki:1994yt,Yamamoto:1996qq,Yamamoto:1994te,Hamazaki:1995dy,Yamamoto:1995tk,GarciaBellido:1995rv,Garriga:1996pg}.
 These deviations trace back to the difference of the state from the Bunch Davies vacuum after the tunnelling event \cite{Yamamoto:1996qq}. They might help to explain the observed lack of power on the largest scales, but a conclusive observation is lacking due to cosmic variance.

It is also possible that our bubble collided with other ones during the slow-roll regime \cite{Dahlen:2008rd}, which may lead to giant circular structure in the CMB, see e.g.~\cite{Aguirre:2009ug,McEwen:2012uk} among others. Such peculiar patterns have not been found beyond the $2\sigma$ level \cite{Kovetz:2010kv}.

It is possible to realise the one-bubble open inflation scenario with a single scalar field \cite{Bucher:1994gb} responsible for both, the tunnelling event and subsequently inflation. However, functional fine-tuning is needed, since slow-roll inflation requires a small field mass, whereas tunnelling via a Coleman-De Luccia (CdL) instanton hinges on a steep, thin  barrier. Further, in light of the many fields present on the landscape, it seems more natural to employ at least two fields, one responsible for tunnelling and a second (perpendicular) one to instigate slow-roll inflation. Naturally, even more fields could be invoked, for instance to end inflation, generate fluctuations via the curvaton mechanism or generate oscillatory signals during the slow-roll regime by means of turns in the trajectory \cite{Chen:2009we,Chen:2009zp,Shiu:2011qw,Sefusatti:2012ye,Pi:2012gf,Chen:2012ge,Noumi:2012vr,Gao:2012uq}. (see \cite{Achucarro:2010jv,Achucarro:2010da,Cespedes:2012hu,Achucarro:2012sm,Achucarro:2012yr,Achucarro:2012fd} for a series of papers developing an effective field theory framework).

Since we are primarily interested in additional observational handles on open inflation, we focus on a simple two-field model (without turn): $\sigma$ denotes the direction in which tunnelling takes place, and a perpendicular field, $\phi$, subsequently drives inflation. A concrete computation of the CdL instandon in such a two field model can be found in \cite{Sugimura:2011tk}.

\subsection{The Setup}

Consider two canonically normalized scalar fields with the action
\begin{equation}
\label{lagrangian}
S=\int d^4x \sqrt{-g}\left(\frac{M_\text{pl}^2}{2} R -\frac{1}{2}g^{\mu\nu}\partial_\mu \sigma \partial _\nu \sigma - \frac{1}{2}g^{\mu\nu}\partial_\mu \phi \partial _\nu \phi -W(\sigma,\phi) \right)\,.
\end{equation}
Given the desired distinction between $\sigma$ and $\phi$, we take a separable potential
\begin{eqnarray}
W=V(\phi)+\frac{1}{2}m_\sigma^2\sigma^2\,,
\end{eqnarray}
where $V(\phi)$ is assumed to be sufficiently flat to yield slow-roll inflation (in this study, we won't need to specify its exact form for the most part). The metric of an open, FRW universe is
\begin{equation}
\label{metric}
 ds^2=-dt^2+a^2\left[\frac{dr^2}{1- k r^2}+r^2\left(d\theta ^2+\sin ^2\theta d\phi ^2\right)\right] , 
\end{equation}
with $k=-1$ and we use the notation ${(\dots)}\dot\, \equiv \partial (\dots)/\partial t$. Occasionally, we use conformal time, $a d\tau=dt$, to simplify analytic expressions.

We treat $\sigma$ as a small, homogeneous perturbation to the background, so that we can define a background Hubble parameter via
\begin{equation}
\label{Friedman}
H_{\text{bgr}}^2\equiv \frac{\rho _{\phi }}{3 M_{\text{pl}}^2}-\frac{k}{a^2}\,,
\end{equation}
which dominates the Hubble friction in the equation of motion of $\sigma$,
\begin{equation}
\label{sigma eom}
\ddot{\sigma }+3 H_{\text{bgr}} \dot{\sigma }+m_{\sigma }^2\sigma \approx 0\,.
\end{equation}
To ease notation, we set the reduced Planck mass to unity throughout  this article,
\begin{eqnarray}
M_\text{pl}=\frac{1}{\sqrt{8\pi G}}\equiv 1\,.
\end{eqnarray}
We note that $\sigma\neq 0$ after tunnelling \cite{Sugimura:2011tk}, but even if it were, it could get displaced due to a feature in the inflationary attractor valley, see Sec.~\ref{sec:excitation} for a brief discussion; we do not explicitly incorporate features in our potential, but use this possibility as a motivation to treat the initial amplitude $\sigma_0$ as a free parameter, set either directly after tunnelling, or at a later time. The maximal oscillation amplitude is bounded from above, if a truncated single-field description is to be used at the perturbed level; we elaborate on this subtle issue in Sec.~\ref{sec:truncating}.  
Any displacement of $\sigma$ leads to damped oscillations and thus a contribution to the Hubble parameter
\begin{equation}
\label{H}
H^2=\frac{\rho_\sigma}{3}+H^2_\text{bgr},
\end{equation}
of  
\begin{equation}
\label{Hsigma}
\rho_\sigma=\frac{1}{2}\left(\dot{\sigma }^2+m_{\sigma }^2\sigma ^2\right)\equiv 3H_\sigma^2\,.
\end{equation}
To treat $\rho_\sigma$ perturbatively, we need 
\begin{eqnarray}
\frac{\rho_\sigma}{\rho_\phi+\rho_k}\ll 1\,,
\end{eqnarray}
where we defined 
\begin{eqnarray}
\rho_k\equiv \frac{-3k}{a^2}=\frac{3}{a^2}\,,
\end{eqnarray}
which leads to an upper bound on $\sigma_0$ that is, however, weaker than the one we impose later on to stay within the regime of a truncated single-field description at the perturbed level, see Sec.~\ref{sec:excitation}. By choosing the normalization $k=-1$, the scale factor has units of $1/\text{mass}$.

Since we are primarily interested in the transition from the curvature dominated regime to inflation, we sometimes approximate
\begin{eqnarray}
\rho_\phi\approx \,\text{const}
\end{eqnarray}
at the background level \footnote{We do not suggest that inflation is driven by a cosmological constant, but merely make this approximation for certain background quantities.}.

We will need the leading order oscillations of the Hubble parameter, \(H_{\text{osci}}\), to compute corrections to correlation functions of the curvature fluctuation $\zeta$; these are caused by resonances with mode functions in the in-in formalism, see Sec.~\ref{sec:powerandbiinin}. To this end we separate the Hubble parameter into a smoothly varying part, and an oscillatory one 
\begin{eqnarray}
H\equiv H_0+H_{\text{osci}} \,,
\end{eqnarray}
where $H_{\text{osci}}\ll H_0$. Since the only oscillatory contribution stems from $\sigma(t)$, we can identify $H_{\text{osci}}$
by comparing (\ref{H}) with
\begin{eqnarray}
 H^2 &\approx &H_0^2+2 H_0 H_\text{osci}\,.
\end{eqnarray}
Since a closed form expression for $H_{\sigma}$ is not  attainable in the transition region to inflation, we apply a moving average to (\ref{Hsigma}) in order to extract the average and the corresponding oscillatory component numerically in this case.
 Given $H_\text{osci}$, the leading order oscillatory contributions to combinations of the slow-roll parameters
\begin{eqnarray}
\epsilon=-\frac{\dot{H}}{H^2}\quad , \quad
\eta=\frac{\dot{\epsilon}}{\epsilon H}
\end{eqnarray}
that we need later on become
\begin{eqnarray}
\epsilon_{\text{osci}}&\approx&-\frac{(\dot{H})_\text{osci}}{H_\text{bgr}^2}\,,\\
(\epsilon\dot{\eta})_\text{osci}&\approx& \frac{1}{H_\text{bgr}^3}\left(-\dddot{H}+\frac{\ddot{H}^2}{\dot{H}}\newline\right.\left.+\frac{3\ddot{H}\dot{H}}{H}-\frac{4\dot{H}^3}{H^2}\right)_\text{osci}\,.
\end{eqnarray}

\subsection{From Curvature Domination to Inflation \label{sec:history}}

We can identify three different regimes in the evolution,
\begin{enumerate}
\item curvature domination, $\rho_k\gg \{\rho_\phi,\rho_\sigma\}$,
\item an intermediate regime (transition to inflation),  $\rho_k \sim \rho_\phi\gg \rho_\sigma$,
\item and inflation, $\rho_\phi \gg \{\rho_k, \rho_\sigma\}$,
\end{enumerate}
each of which we can treat separately. 
Since simple analytic expressions are often attainable in conformal time,  we shall freely switch between $t$ and $\tau$ in the following. We put the tunnelling event in the far past, $\tau_\text{tun}\ll 0$. 

Since spatial curvature dominates the energy balance initially, the scale factor grows exponentially with $\tau$. We start our investigation at
\begin{eqnarray}
a= a_0^{(k)}=1\,,
\end{eqnarray} 
so that the curvature contribution to the Friedmann equation is unity in natural units. This phase ends once the curvature contribution approaches that of the inflaton at
\begin{eqnarray}
a^{(k)}_\text{f}\equiv\frac{1}{A \sqrt{\rho_\phi/3}}\,,
\end{eqnarray}
with $A\gtrsim 1$ (we choose $A\equiv 10$).  As the expansion progresses, energy stored in  spatial curvature continues to red-shift  $\propto a^{-2}$, while $\rho_\phi\approx \text{const.}$ and $\rho_\sigma\propto a^{-3}$ ($\sigma$ oscillates in a quadratic potential). After a brief intermediate regime, inflation commences, during which we approximate $a\propto t^p$ with $p\gg 1$. We define
 \begin{eqnarray}
a_\text{i}^\phi\equiv \frac{1}{B \sqrt{\rho_\phi/3}}\label{aiphi}\,,
 \end{eqnarray}
  with $B\lesssim 1$ so that $\phi$ dominates for $a>a_\text{i}^{\phi}$ (we choose $B\equiv 1/A=0.1$). We denote the instance of equality by $\tau_\text{eq}$, when $H_\phi^2=H_\text{k}^2$, or equivalently 
  \begin{eqnarray}
  a=a_\text{eq}\equiv \frac{1}{\sqrt{\rho_\phi /3}} \,.
  \end{eqnarray}
We  count the number of e-folds positive from here on,
\begin{eqnarray}
N\equiv \int_{t_\text{eq}}^{t_\text{end}} H\, dt\,,
\end{eqnarray}
and assume a $V(\phi)$ such that inflation ends once 
$N\gtrsim 60$ and the remaining curvature is diluted  sufficiently to solve the flatness problem. The length of slow-roll inflation can be tuned freely by adjusting $V(\phi)$ appropriately. At the end of inflation, any $\sigma$-oscillations have died away, since $\sigma$  is assumed to be considerably heavier than  the inflation.  During inflation, the leading order term to the slow-roll parameter is constant and small, $\epsilon=p^{-1}$.

Given these three regimes, it is simple to derive approximate analytic expressions at the background level.
A particular combination that enters our subsequent computations is defined by
\begin{equation}
\label{f}
f(\tau)\equiv \frac{1}{a}\frac{d a}{d\tau}+\frac{1}{2\epsilon_\text{bgr}}\frac{d \epsilon_\text{bgr}}{d \tau}\,,
\end{equation}
and  a straightforward computation, entailing approximate solutions to the Friedmann equation, yields
\begin{equation}
f(\tau)\approx 
\begin{cases}
\frac{a_0^{(k)}}{t_0} & \text{for}\, 1=a_0^{(k)}<a< a_f^{(k)}, \:\: \text{curvature domination}\\
-\tanh(\tau)& \text{for}\, a_f^{(k)} \leq a \leq  a_i^\phi, \: \: \text{intermediate regime}\\
\frac{p}{(1-p)\tau} & \text{for}\, a > a_i^\phi,  \:\: \text{inflation}.\\
\end{cases}
\end{equation}
The $t_0$ corresponds to the choice of initial time, see Sec.~\ref{sec:curv dom}. 

\section{Resonant Signals in the In-In Formalism}
We want to compute correlation functions of the gauge invariant curvature perturbation $\zeta$, which are observationally accessible in surveys of the cosmic microwave background radiation (CMBR) and large scale structure (LSS). $\zeta$ is a popular choice, since it is conserved in single-field models once a mode crosses the horizon. However, if more fields are present, any turn in the trajectory couples perturbations perpendicular (isocurvature) and parallel (adiabatic) to the trajectory \cite{Gordon:2000hv}, which in turn can alter predictions \cite{Shiu:2011qw}.

In our setup, we are interested in oscillations of $\sigma$, which can be the result of a displacement directly after the tunnelling event or a subsequent event such as a drop or a sharp turn in the trajectory (by sharp, we mean $1/\Delta t_\text{turn}>m_\sigma$, in which case damped oscillations with frequency $m_\sigma$ result \cite{Gao:2012uq}).

There are two effects that can give oscillatory signatures in correlation functions of $\zeta$:
\begin{enumerate}
\item The \emph{direct coupling} between perturbations in the fields, which is governed by the rate of change in the direction of the trajectory, see e.g.~\cite{Achucarro:2010jv,Shiu:2011qw,Gao:2012uq}.
\item The \emph{modulation of the background} brought forth by $\sigma(t)$, see e.g.~\cite{Chen:2010xka,Chen:2011zf,Chen:2011tu,Chen:2012ja}. 
\end{enumerate}

The first effect has its origin in multi-field dynamics, that requires a careful investigation of perturbations in all fields. To this end, the effective field theory (EFT) of single-field inflation in the presence of heavier fields has been developed (not to be confused with the EFT of single-field inflation \cite{Cheung:2007st,Weinberg:2008hq,Senatore:2009cf}), to account for models that are consistent with the rational of a low energy effective theory \cite{Achucarro:2010jv,Achucarro:2010da,Cespedes:2012hu,Achucarro:2012sm,Achucarro:2012yr,Achucarro:2012fd}, see also \cite{Shiu:2011qw,Burgess:2012dz} (some concerns regarding the applicability of an EFT raised in \cite{Avgoustidis:2012yc} were primarily due to different definitions of the term EFT, see e.g.~the discussion in \cite{Burgess:2012dz}). The resulting action for perturbations is that of a single scalar field with a time dependent speed of sound $c_s(t)$, which is set by the background evolution. While a kinematic basis for fluctuations \cite{Achucarro:2010jv,Achucarro:2010da,Cespedes:2012hu,Achucarro:2012sm,Achucarro:2012yr,Achucarro:2012fd} is appropriate 
for a slow turn (see also \cite{Chen:2009we,Chen:2009zp,Chen:2012ge}), a mass basis is better suited for sudden turns \cite{Gao:2012uq}. However, the effective field theory tends to break down easily for sudden turns, requiring a true multi-field analysis at the perturbed level. Considerable progress has been made in the last years to understand signals in the late time adiabatic mode caused by such multi-field dynamics. However, in most cases studied analytically, the background evolution is approximated to be smooth, i.e.~$H\approx \mbox{const}$ is assumed during the turn. 

The second effect takes into account the change in background quantities induced by oscillations of heavy fields, leading to additional oscillatory components to e.g.~$H(t)$ or slow-roll parameters \cite{Chen:2010xka,Chen:2011zf,Chen:2011tu,Chen:2012ja}. At the perturbed level, a truncation to the light field is performed, that is, any interactions with perturbation in the heavy field(s) are ignored. It is then relatively straightforward  to use the in-in formalism to compute additional features in correlation functions caused by resonances between mode functions and oscillations of background quantities.

In a realistic scenario both effects are present. Unfortunately, to our knowledge, the added complexity of the problem in presence of an oscillating background and true multi-field dynamics for perturbations has so far prevented a comprehensive investigation of observational signals (see however recent advances such as the case study in \cite{Collins:2012nq}). In the meantime, we may focus on either of the two effects to estimate the feasibility of creating detectable signals -- if it turns out that signals are unobservable for the second effect, one may confidently constrain oneself to the first one (and vice versa). However, if both effects lead to observably large signals, one should follow up with a study combining them, before comparing predictions with data.

In this paper we focus on the second effect, that is we truncate perturbations to the inflaton, while ignoring perturbations in the tunnelling field. To apply this logic, any coupling of perturbations in the heavy field, $\delta\sigma$, to the fluctuations in the light one, $\delta\phi$, have to be much smaller than existing terms in the equations of motion for $\delta\phi$ -- otherwise the treatment is inconsistent. The absence of concrete viability criteria for the truncation in the work by Chen et.al.~ \cite{Chen:2010xka,Chen:2011zf,Chen:2011tu,Chen:2012ja} led to some lively discussions. Hence, we would like to elaborate briefly on the conditions under which such a truncation is viable.

\subsection{Truncating Perturbations to the Light Sector \label{sec:truncating}}

Consider the gauge invariant field perturbation
\begin{eqnarray}
Q_I\equiv \delta\phi_I+\frac{\dot{\phi}_I}{H}\psi\,,
\end{eqnarray} 
where $I\in\{\phi,\sigma\}$, $\phi_\phi=\phi,\phi_\sigma=\sigma$, and $\psi$ denotes the $ii$ perturbation of the metric \cite{Mukhanov:1990me}. The equations of motion for $Q_I$ are of the form \cite{Gordon:2000hv}
\begin{eqnarray}
\ddot{Q}_I+3H\dot{Q}_I+\frac{k^2}{a^2}Q_I+\sum_J M_{IJ}Q_J=0\,,
\end{eqnarray} 
where the mass matrix entails the square of the fields bare masses as well as simple combinations of the  background solutions. Diagonalization of the potential's mass matrix $W_{IJ}\equiv \partial^2 W/\partial \phi_I\partial \phi_I$ yields the mass basis of the fields in the potential $W(\phi,\sigma)$; for us, this basis is given directly by $\sigma$ and $\phi$, since we consider a separable potential. To be concrete, we take a quadratic potential for the inflaton in this section
\begin{eqnarray}
 V(\phi)\equiv\frac{1}{2} m_\phi^2\phi^2\,;
 \end{eqnarray}
a generalization to other potentials is straightforward. 

Directly after tunnelling, perturbations in both fields should be of the same order, $Q_\sigma\sim Q_\phi$, and since $Q_\sigma$ decays faster subsequently due to $m_\sigma\gg m_\phi$, we can safely take $Q_\sigma\lesssim Q_\phi$. Hence, to self-consistently ignore $Q_\sigma$, we need  $M_{\phi\phi}\gg M_{\phi\sigma}$ as a necessary condition. To compare terms in the mass matrix, we take $H^2\approx H_{\text{bgr}}^2$, assume slow-roll for the inflaton and damped oscillations for $\sigma$ with initial amplitude $\sigma_0$. Since perturbations in the inflaton need to satisfy the COBE bound, we take $m_\phi\sim \sqrt{8\pi}\times 10^{-6}$, while sixty e-folds of inflation requires $\phi\sim 16 $ at the onset of inflation. Thus, to estimate the order of magnitude of terms, we take
\begin{eqnarray}
|\dot{\phi}|&\sim& \frac{m_\phi^2\phi}{H}\quad\,,\quad
|\ddot{\phi}|\ll |H\dot{\phi}|\,,\label{est1}\\
|\sigma| &\sim& \sigma_0 \quad \,, \quad |\dot{\sigma}| \sim \sigma_0 m_\sigma\quad\,,\quad|\ddot{\sigma}|\sim \sigma_0 m_\sigma^2\,.\label{est2}
\end{eqnarray} 
We further define
\begin{eqnarray}
\delta^2\equiv \frac{\rho_\phi}{3H^2}\approx \frac{V(\phi)}{3H^2}\leq 1\,,\label{defdelta}
\end{eqnarray}
which approaches one towards the inflationary regime. To be precise, we have $\delta(a_\text{f}^{(k)})=1/\sqrt{A^2+1}\approx 0.1$ at the end of the curvature dominated regime, $\delta(a_\text{eq})=1/\sqrt{2}$ at equality and $\delta(a_\text{i}^\phi)=1/\sqrt{B^2+1}\approx 0.995$ at the beginning of the inflationary regime.   We first note that 
\begin{eqnarray}
M_{\phi\phi}\approx m_\phi^2+\mathcal{O}
\left(\frac{\dot\phi\ddot{\phi}}{H}\,,\, 
 \dot{\phi}^2\,,\,  
 \frac{\dot{\phi}^2\dot{H}}{H^2}\right)
\end{eqnarray}
 to leading order in slow-roll. The contributions to the mixed mass matrix elements are of order
 \begin{eqnarray}
M_{\phi\sigma}\sim\mathcal{O}\left(\dot{\phi}\dot{\sigma}\,,\,\dot{\phi}\dot{\sigma}\frac{\dot{H}}{H^2}\,,\,\frac{\dot{\phi}\ddot{\sigma}}{H}\,,\,
\frac{\ddot{\phi}\dot{\sigma}}{H}\right) \,.
 \end{eqnarray}
 During the curvature dominated phase $\dot{H}/H^2\rightarrow -1$, while it is slow-roll suppressed during inflation, so that the second term on the r.h.s.~is at most of the same order as the first one. Further, since $|\ddot{\phi}|\ll |H\dot{\phi}|$ due to slow-roll, we see that the fourth term is smaller than the first one and we only need to compare the first and third term with $m_\phi^2$. Writing $H\sim \phi m_\phi/\delta$ and using (\ref{est1}) as well as (\ref{est2}), we arrive at the conditions 
 \begin{eqnarray}
 \label{constraintsigma0a}
\sigma_0&\ll&\frac{m_\phi}{m_\sigma}\frac{1}{\delta}\,,\\
  \label{constraintsigma0b}
  \sigma_0&\ll&\frac{m_\phi^2}{m_\sigma^2}\phi\frac{1}{\delta^2}\,.
 \end{eqnarray}
Thus, the heavier the tunnelling field, the lower the allowed amplitude of oscillations in $\sigma$ if we want to neglect the effect of $\delta \sigma$ on $\delta \phi$. The bounds are strongest in the inflationary regime, but get relaxed during curvature domination, since $\delta\ll 1$. If $V(\phi)$ is more complicated, similar bounds result; we will resort to the above quadratic potential whenever the need for numbers arises. 

\subsubsection{Exciting Heavy Fields\label{sec:excitation}}

The heavy field $\sigma$ is usually displaced from the inflationary valley directly after tunnelling \cite{Sugimura:2011tk} \footnote{This initial excitation is thus similar to the initial excitation of heavy fields considered in \cite{Burgess:2002ub}.}, but it can also get excited subsequently, for example by encountering a feature in the potential leading to a turn as in \cite{Shiu:2011qw}; it could also run into a sharp drop in the $\sigma$ direction, which might be more appealing: consider that after tunnelling, $\sigma$ finds itself on a reasonably flat plateau of the potential, moving slowly due to large Hubble friction; after some time, but still during curvature domination, $\sigma$ encounters a sudden drop, leading for instance into the quadratic well with $m_\sigma\gg m_\phi$ that was used above; the dynamics are thus reminiscent to the ones of the waterfall-field at the end of hybrid inflation. We expect that the excitation event has a strong effect on the perturbations of the involved field(s). However, if only the $\sigma$-dynamics are affected, e.g.~due to a drop, fluctuations in $\phi$ are oblivious to the excitation event itself,
 as long as the conditions in (\ref{constraintsigma0a}) and (\ref{constraintsigma0b}) are satisfied and $Q_\sigma$ is not amplified above $Q_\phi$. A turn in the trajectory is more problematic than a drop, since it actively mixes up the roles of the fields, requiring a careful multi-field treatment at the perturbed level \cite{Shiu:2011qw}. Hence, we take as a working hypothesis that the excitation event is of the former type, so that fluctuations in $\phi$ are unaltered by the excitation of $\sigma$. It would be interesting to investigate such a drop numerically, which goes however beyond the scope of the current article.

There is a further constraint we need to impose though: since we treat $\rho_\sigma$ as a small correction to the background energy, we need to demand $\rho_\sigma \ll \rho_k$ during the curvature dominated regime and $\rho_\sigma\ll \dot{\phi}^2/2$ during inflation\footnote{The kinetic energy, not the potential one, matters here, since the former determines  $\epsilon=-\dot{H}/H^2$ which should remain  largely unaffected by any sudden changes in $\rho_\sigma$ and its derivatives during the excitation (for example, if kinetic energy in $\phi$ is used to excite $\sigma$, whose potential energy red-shifts as $a^{-3}$, $\dot{H}$ increases temporarily); such a sudden change can lead to considerable particle production in  $Q_\phi$, with an accompanying bump and ringing in the power-spectrum \cite{Battefeld:2010rf} (similar to the signals in \cite{Joy:2007na} caused by a jump in the mass of the inflaton), in addition to the signals computed in this paper (see also \cite{Shiu:2011qw}).}. Using quadratic potentials for both fields as before, these conditions entail
\begin{eqnarray}
\sigma_0\ll \frac{m_\phi}{m_\sigma}\frac{\phi}{\delta}
\end{eqnarray}
during curvature domination and
\begin{eqnarray}
\sigma_0\ll\frac{2}{3}\frac{m_\phi}{m_\sigma}
\end{eqnarray}
during inflation, which are comparable  to the constraints in (\ref{constraintsigma0a}) and (\ref{constraintsigma0b}). We enforce these constraints throughout and, keeping the above caveats in mind, treat  $\sigma_0$ and its excitation time as free parameters. 

\subsection{The Curvature Perturbation, Correlation Functions and Mode Functions \label{sec:bd vs non-bd}} 

After having justified a truncation to perturbations in the inflaton, we would like to introduce and comment on quantities of interest, particularly correlation functions of the gauge invariant curvature perturbation $\zeta$,
\begin{eqnarray}
\left<\zeta^n\right> \equiv \left<0\right|\zeta (\textbf{k}_1,0)\dots \zeta (\textbf{k}_n,0) 
\left|0\right>\,.
\end{eqnarray}
The two point function defines  the power-spectrum $\mathcal{P}_\zeta$
\begin{equation}
\label{def power-spectrum}
\left<\zeta^2\right> \equiv \frac{P_\zeta}{2 k_1^3} (2 \pi)^5 \delta(\textbf{k}_1+\textbf{k}_2),
\end{equation}
which, in single-field slow-roll inflation, is
 \begin{equation}
 \label{one field slow-roll power-spectrum}
P_\zeta=\frac{H^2}{8 \pi ^2\epsilon}\,.
 \end{equation}
The three-point function defines the bi-spectrum
\begin{equation}
\label{def shape function}
\left<\zeta^3\right>\equiv S(k_1, k_2, k_3) \frac{P_{\zeta}^2}{\prod\limits_{i} k_i^2} (2 \pi)^7 \delta^3 (\textbf{k}_1+\textbf{k}_2+\textbf{k}_3),
\end{equation}
which is commonly characterised by a dimensionless shape function $S$; we use the conventions of \cite{Chen:2010xka} for ease of comparison, see App.~\ref{app:A}. Translational invariance requires that the three momentum vectors form a triangle, and rotational invariance implies that only the amplitudes of the momenta enter  ($k_i=|\textbf{k}_i|$). The shape of $S$ refers to the dependence on the shape of the triangle formed by the three momenta (i.e.~the dependence on $k_1/k_3$ and $k_2/k_3$ while keeping $K=k_1+k_2+k_3$ fixed); the running of $S$ refers the dependence of $S$ on $K$ for a fixed shape.

What determines $\zeta$?
The equation of motion for adiabatic perturbations can, via field redefinitions, always be rewritten as a simple harmonic oscillator with a time dependent mass (set by the background evolution), see for example the review \cite{Mukhanov:1990me}. The variable satisfying this equation is commonly referred to as the Mukhanov variable $v_k$.

On small scales ($k\gg aH$), the mass term becomes negligible compared to the gradient term $k^2v_k$, and simple oscillatory solutions result for $v_k$. Hence, the presence of spatial curvature is irrelevant for the evolution of $v_k$ on small sales. The amplitudes of the two independent oscillatory solutions are usually set (after quantization) by imposing the Bunch-Davies vacuum state.  

Not surprisingly, the oscillations in $v_k$ are directly related to oscillations in $\zeta$: if we define the mode function of $\zeta$ by
\begin{equation}
\label{mode function def}
u_k=\int d^3 x\, \zeta(t,\textbf{x})e^{-i \textbf{k}\cdot\textbf{x}}\,,
\end{equation}
it is easy to show that a Bunch-Davies state for $v_k$ translates to 
\begin{equation}
\label{mode function}
u_k\approx 
\frac{1}{a\sqrt{4 \epsilon_\text{bgr} k}}e^{-i k \tau }\,.
\end{equation}
These oscillations on small scales are of particular interest to us, since they can resonate with oscillations in background quantities. In the above, we expanded the full solution in the limit  $k\gg aH$, which is a good approximation for all cases discussed in this article.
Once scales cross the horizon, one can show that $\zeta$ is frozen;  subsequently, $\zeta$ sets the amplitude of temperature fluctuations in the CMBR, enabling a direct probe of early universe physics.

Before we compute correlation functions of $\zeta$, we need to address a crucial point: the initial state of fluctuations after the tunnelling event in open inflation is not the Bunch Davies state; a detailed discussion of the state for single and two-field one-bubble open inflation can be found in \cite{Yamamoto:1996qq}. 

The deviation from the standard Bunch-Davies state arises due to the mass difference outside the bubble wall and is parametrised by $1-Y$ with
\begin{equation}
Y=\frac{\Gamma(2-i p)}{\Gamma(2+i p)}\frac{e^{-\pi p} \beta_p}{2 \cosh \pi p\, \bar{\alpha}_p}+\text{c.c.}\,.
\end{equation}
Here $\bar{\alpha}_p$ and $\beta_p$ are coefficients relating the mode function of the true vacuum inside the bubble to the mode functions of the false vacuum outside the bubble, and $p$ labels momenta (for more details, see \cite{Yamamoto:1996qq}). The key observation to be taken away is that on scales smaller than the curvature scale (sub-horizon modes with large $p$), $Y$ tends to $0$, and the state can be approximated by the Bunch-Davies one. This agrees with our intuition (and the equivalence principle) that given a small enough patch, spacetime inside the bubble approaches Minkowski. 

Hence, we proceed with the Bunch-Davies state as long as we are concerned only with modes inside the horizon. We would like to emphasize that any (small) deviation from the Bunch-Davies state inside the Horizon might slightly affect the amplitude and frequency of $u_k$, but not the presence of high frequency oscillations that are key to our upcoming analysis.

\subsection{Overview of the In-In formalism \label{sec:overvieinin}}

The in-in formalism was initially introduced by Schwinger \cite{Schwinger:1960qe} and used more recently in a cosmological setting by Maldacena \cite{Maldacena:2002vr} and  Weinberg \cite{Weinberg:2005vy} among others (for recent reviews see \cite{Baumann:2009ds} and \cite{Adshead:2009cb}). 
It can be used to compute correlation functions of the curvature perturbation $\zeta$, for example the power-spectrum and the shape function of the bi-spectrum. Here, we provide a brief summary following \cite{Chen:2010xka}, primarily to introduce our notation.

Starting from a Lagrangian, one may perturb it around a homogeneous solution of the classical equations of motion; quadratic and higher order terms provide the dynamics of perturbations. Conjugate momenta are defined from the interaction Lagrangian, enabling the computation of the Hamiltonian in the perturbations.  By gathering all terms higher than 2'nd order, one arrives at the interaction Hamiltonian $H_I$. Solutions of the classical equations of motion for the quadratic part of the Hamiltonian in Fourier space are commonly called mode functions, which are normalized according to a Wronskian (quantization) condition with initial conditions corresponding (usually) to the Bunch-Davies vacuum. A correlation function is then computed as the nested integral expression
\begin{equation}
\label{ininintegral}
\left\langle \zeta ^n (t) \right\rangle =\left\langle 0\left|\left[ \bar{T}\exp \left( i \int\limits_{t_0}^{t}  H^I d t  \right)\right]\zeta^n(t)\left[T\exp \left(-i \int\limits_{t_0}^{t} H^I d t  \right)\right]\right|0 \right\rangle\,,
\end{equation}
where $T$ is the  time-ordering operator ($\bar{T}$ the reverse one). After expanding $H$ to the desired order, one can perform contractions, sum over them and perform a time-ordered integration.

Our aim is to compute the two and three-point correlation function of $\zeta$.  The Hamiltonian $H$ (not to be confused with the Hubble parameter, which does not play a role in this section) is a functional of the field $\phi$ and its conjugate momentum $\pi$. It determines the system's evolution via
\begin{equation}
\dot{\phi}(\textbf{x},t)=i\left[H(\phi(t),\pi (t)),\phi(\textbf{x},t)\right]\qquad , \qquad \dot{\pi}(\textbf{x},t)=i\left[H(\phi(t),\pi (t)),\pi(\textbf{x},t)\right]\,.
\end{equation}
We expand the field in terms of a time-dependent background (denoted by a bar) and perturbations, i.e.~$\phi(\textbf{x},t)=\bar{\phi}(t)+\delta \phi (\textbf{x},t)$. The background evolves according to the classical equations of motion,
\begin{equation}
\dot{\bar{\phi}}(t)=\frac{\partial \mathcal{H}}{\partial \bar{\pi}}, \qquad \dot{\bar{\pi}}(t)=-\frac{\partial \mathcal{H}}{\partial \bar{\phi}},
\end{equation}
where $\mathcal{H}$ is the Hamiltonian density. Since the background commutes with everything (it is just a complex number), the canonical commutation relations are transferred to the perturbations alone,
\begin{eqnarray}
\nonumber &&\left[\delta \phi (\textbf{x},t), \delta \pi (\textbf{y},t)\right] = i \delta^3(\textbf{x}- \textbf{y}),\\
&&\left[\delta \phi (\textbf{x},t), \delta \phi (\textbf{y},t)\right] =\left[\delta \pi (\textbf{x},t), \delta \pi (\textbf{y},t)\right] = 0.
\end{eqnarray}
Taylor expanding the Hamiltonian around the background solution yields
\begin{eqnarray}
\nonumber H\left[\phi(t), \pi (t)\right]&=&H\left[\bar{\phi}(t), \bar{\pi} (t)\right]+ \int d^3x \frac{\partial \mathcal{H}}{\partial \bar{\phi}(t)}\delta \phi(\textbf{x},t)+\int d^3x \frac{\partial \mathcal{H}}{\partial \bar{\pi}(t)}\delta \pi	(\textbf{x},t)\\&&+ H_0\left(\delta \phi (t),\delta \pi (t),t\right)+H_I\left(\delta \phi (t),\delta \pi (t),t\right).
\end{eqnarray}
Here, $H_0$ contains quadratic terms in the field, which determine the leading evolution of the perturbations, and $H_I$ contains all higher orders. Combining these results, we see that the equations of motion for the perturbations are determined by $H_0+H_I$. The point of separating the non-linear Hamiltonian in two components is to determine first the kinematic evolution as driven by $H_0$, and then perturbatively compute the effects of interactions induced by $H_I$. For a generic operator $Q$ in a state $\left| \Omega\right\rangle$, the expectation value of $Q$ is \citep{Chen:2010xka}
\begin{equation}
\label{general ininintegral}
\left\langle \Omega\right | Q\left(\delta \phi (\textbf{x},t),\delta \pi (\textbf{x},t)\right)\left | \Omega \right \rangle=
\left\langle \Omega\right |F^{-1}(t,t_0) Q\left(\delta \phi^I (\textbf{x},t),\delta \pi^I (\textbf{x},t)\right)F(t,t_0)\left | \Omega \right \rangle
\end{equation}
with $t_0$ in the far past and 
\begin{equation}
F(t, t_0)=T\exp \left(-i \int\limits_{t_0}^{t}H_I (t)dt\right).
\end{equation}
Hence, for $Q=\zeta$ (\ref{general ininintegral}) reduces to (\ref{ininintegral}).

Note that after quantisation, the introduction of mode functions, and application of $T$ and $\bar{T}$, the creation and annihilation operators kill the respective left and right vacuum states. Hence, the remaining terms are those where the fields are contracted away. Further, one can use Feynman diagrams to keep track of the necessary contractions, while keeping in mind that the proper vacuum state has to be prescribed in order to compute an explicit integral.

\subsection{The Power- and Bi-spectrum \label{sec:powerandbiinin}}

To leading order in $\Delta \epsilon=\epsilon_{osci}$ and $\Delta (\epsilon\dot{\eta})=(\epsilon\dot{\eta})_{osci}$, the interaction Hamiltonian can be expanded to \citep{Chen:2011zf}\footnote{These results were derived in a spatially flat universe, but they remain valid in the regime of interest if curvature is present.}
\begin{eqnarray}
\mathcal{H}_0&=&a^3 \epsilon_0 \dot{\zeta}^2-a \epsilon_0 (\partial\zeta)^2\,,\\
\label{H2}
\mathcal{H}_2^I&\approx&-a^3 \Delta \epsilon \dot{\zeta}^2+a \Delta \epsilon (\partial\zeta)^2\,,\\
\label{H3}
\mathcal{H}_3^I&\approx&-\frac{1}{2}a^3 \Delta(\epsilon \dot{\eta}) \zeta^2 \dot{\zeta}+\dots \,.
\end{eqnarray}
We only  kept the most important term in $\mathcal{H}_3^I$ (the one with the most time derivatives acting on slow-roll parameters, as in \cite{Chen:2006xjb,Chen:2008wn}). Thus, the leading correction to the two-point correlation function is given by 
\begin{equation}
\Delta\left\langle \zeta^2\right\rangle=\left\langle 0\right| i\int\limits_{\tau_\text{min}}^{\tau_\text{max}}a d\tau d^3 x\left[\mathcal{H}_2^I, \zeta_{\textbf{k}_1}\zeta_{\textbf{k}_2}\right]\left|0\right\rangle \,.
\end{equation}

Using the definition of the power-spectrum in (\ref{def power-spectrum}) and the unperturbed power-spectrum of a single-field  slow-roll model in (\ref{one field slow-roll power-spectrum}), one arrives at 
\begin{equation} 
\label{def power-spectrum correction}
\frac{\Delta P_\zeta}{P_{\zeta}}(k)=2i\int _{\tau_\text{min}}^{\tau_\text{max}}d\tau  a^2\Delta \epsilon  \left(\left.u_k^\backprime
\right.^2-k^2u_k^2\right)+c.c. \,.
\end{equation}
Similarly, the leading order contribution to the 3-point correlation function is
\begin{eqnarray}
\nonumber \left< \zeta^3\right>
&=&\left< 0\right | i \int\limits_{\tau_\text{min}}^{\tau_\text{max}} a d \tau d^3 x \left[\mathcal{H}_3^I, \zeta^3(\textbf{k}_1 \textbf{k}_2 \textbf{k}_3)\right] \left|0\right>\\
\nonumber& =& i (2 \pi)^3 \delta^3(\textbf{k}_1+\textbf{k}_2+\textbf{k}_3) \left(\prod\limits_{i} u_{k_i}(0)\right)
\int\limits_{\tau_\text{min}}^{\tau_\text{max}} d\tau a^3 \Delta(\epsilon \dot{\eta}) u_{k_1}^*u_{k_2}^*\frac{d u_{k_3}^*}{d \tau}\\
&&+ 2 \text{ permutations} + c.c.\,,
\end{eqnarray}
so that the shape function in (\ref{def shape function}) becomes
\begin{equation}
\label{shape function}
S\left(k_1,k_2,k_3\right)=\frac{(k_1 k_2 k_3)^2}{(2 \pi)^4 P_{\zeta}^2} i\int _{\tau_\text{min}}^{\tau_\text{max}}d\tau  a^3\Delta(\epsilon  \dot{\eta })u_{k_1}^*u_{k_2}^*\frac{du_{k_3}^*}{d\tau }+\text{2 perm.} + c.c.\,.
\end{equation}
Analytic approximations for (\ref{def power-spectrum correction}) and (\ref{shape function}) in inflationary regimes are given in \citep{Chen:2011zf}, which we recover (and slightly extend) in section \ref{sec:infl dom}. The corresponding results in the presence of curvature are provided in Sec.~\ref{sec:curv dom} and Sec.~\ref{sec:com contr}.

Substituting the mode function into (\ref{def power-spectrum correction}) and (\ref{shape function}), and incorporating the complex conjugates, the expressions simplify to
\begin{equation}
\label{power simple}
\frac{\Delta P_\zeta}{P_{\zeta}}(k)= 
\int _{\tau_\text{min}}^{\tau_\text{max}}\left(\frac{f(\tau)^2 -2k^2}{k} \sin(2 k \tau)+2 f(\tau)\cos(2 k \tau)\right) \frac{\Delta\epsilon}{\epsilon_\text{bgr}}d\tau
\end{equation}
 and
\begin{equation}
\label{shape simple}
S\left(k_1,k_2,k_3\right)=k_1 k_2 k_3\frac{1}{(a^3 H_\text{bgr}^4)|_{(t_0)}}
\frac{\sqrt{\epsilon_\text{bgr}}}{16}
 \int _{\tau_\text{min}}^{\tau_\text{max}}
 \frac{\Delta(\epsilon \dot{\eta})}{\epsilon_\text{bgr}^{3/2}}
\left(-2 K\cos(K\tau)+ 6 f(\tau)\sin(K \tau) \right) \,d\tau \,,
\end{equation}
with $f(\tau)$ from (\ref{f}), encoding the time-dependent background evolution of the amplitude of $u_k$, and we defined rescaled wave-numbers
\begin{eqnarray}
\tilde{k_i}\equiv \frac{k_i}{k_\text{eq}}\quad , \quad \tilde{k}\equiv  \frac{k}{k_\text{eq}}\quad , \quad
\tilde{K}\equiv  \frac{K}{k_\text{eq}}= \frac{k_1+k_2+k_3}{k_\text{eq}}\,,
\end{eqnarray}
where 
\begin{eqnarray}
k_\text{eq}=a(t_\text{eq})H(t_\text{eq})
\end{eqnarray}
is the wave-number corresponding to the Hubble horizon at equality\footnote{In all plots, we choose appropriate approximations for $H$ in the three regimes, e.g.~during the curvature dominated regime we use $H\approx \sqrt{\rho_k/3}$.}, $\rho_k=\rho_\phi$;  
for ease of notation, we dropped the tilde
\begin{eqnarray}
\tilde{k}_i\rightarrow k_i\quad,\quad \text{etc.}
\end{eqnarray}
In the remainder of this article, all wave-numbers are normalized in this manner. 

The terms containing $f(\tau)$ in (\ref{shape simple}) and (\ref{power simple}) are missing in \cite{Chen:2011zf}, since they are sub-leading during inflation, but they play an important role if curvature is important. The pre-factor  $k_1 k_2 k_3$ in (\ref{shape simple}) is a due to the chosen normalization, see Appendix \ref{app:A}; thus, we plot $S/(k_1k_2k_3)$ in figures. 

The integration limits above must be adjusted to the boundaries of the region under consideration -- this is crucial for the transmission of a signal, see Sec.~\ref{sec:curv dom}, \ref{sec:infl dom} and \ref{sec:com contr}. As discussed in section \ref{sec:bd vs non-bd}, we take the initial state to be the Bunch-Davies vacuum.
In order to compute the integrals (\ref{power simple}) and (\ref{shape simple}), we note the high-frequency oscillations of the mode function (\ref{mode function}) on sub-horizon scales and the oscillations of the slow-roll parameters induced by the massive $\sigma $-field, whose frequency increases with the mass of the field. They allow the leading order behaviour of the integrals to be analytically approximated via the stationary phase approximation. For details on this asymptotic method, see e.g.~\cite{Murray}, chapter 4. The relevant formula is 
\begin{equation}
\int_{-\infty}^{\infty} \psi(z)e^{i F(z)} dz\sim \psi(z_*)e^{i(F(z_*)+\text{sign}(F''(z_*))\frac{\pi}{4})}\sqrt{2 \pi/|F''(z_*)|},
\end{equation}
for $F(z)\rightarrow \infty$ and $\psi(z)$ sufficiently well-behaved (smooth and without high-frequency oscillations) as well as $F'(z_*)=0$. Away from this equilibrium point, the integrand oscillates rapidly so that the contribution to the integral cancels out approximately over one period. If the equilibrium point falls within the limits of integration in one of our integrals, a signal is generated; if it falls outside, the contribution vanishes. This asymptotic approximation  introduces a step function at the boundary, i.e.~when the equilibrium point coincides with a limit of integration, so that no remnants of a signal can seep through it; in reality, a continuous transition occurs. Since we rely on this analytic approximation, we expect to obtain results with sharp boundaries and possible discontinuities.

We focus on analytical approximations in order to gain insight into the resonance mechanism. Furthermore, due to the highly oscillatory integrand, a numerical computation is challenging and usually of limited practical use, since no direct detection of oscillations in the power or bi-spectrum has been made yet.

\section{Oscillatory Signals in Correlation Functions}

We have now the tools at our disposal to compute the signals in the power-spectrum (\ref{power simple}) and bi-spectrum (\ref{shape simple}) brought forth by the oscillating tunneling field $\sigma$. We discuss the three regimes introduced in Sec.~\ref{sec:history} consecutively: curvature domination in Sec.~\ref{sec:curv dom}, inflation in Sec.~\ref{sec:infl dom} and the intermediate regime in Sec.~\ref{sec:com contr}.

\subsection{Curvature Domination \label{sec:curv dom}}

In a curvature dominated universe we can ignore the inflaton field, and solve the Friedmann equation (\ref{Friedman}) to
\begin{equation}
\frac{a}{a_0^{(k)}}=\frac{t}{t_0}.
\end{equation}
We start our discussion of the curvature dominated regime at $a(t_0)=a_0^{(k)}=1$ with $t_0=1$ so that $\,a=t$. Further, for computational convenience, we work with conformal time $\tau$, normalised to the time of equality, 
\begin{equation}
\tau_0=-\frac{t_0}{a_0^{(k)}}\ln\left(\frac{t_\text{eq}}{t_0}\right)\,.
\end{equation}
 In other words, we convert the entire region of interest (plus a neighbourhood) to the negative real axis via the map $t\mapsto\tau$ : $(0, t_\text{eq})\mapsto (-\infty,0)$. This leads to the following relations for the conformal time\footnote{Thus, conformal time is identical to the number of e-folds normalised to $a_\text{eq}$, $\tau=N\equiv\ln \left(a/a_\text{eq}\right)$.}, the scale factor, the background Hubble parameter, and the background slow-roll parameter 
\begin{eqnarray}
\tau =\ln \left(\frac{t}{t_\text{eq}}\right)\quad,\quad 
 a(\tau )=\sqrt{\rho_\phi/3}e^\tau\quad ,\quad 
H_{\text{bgr}}(t)=\frac{1}{t}\quad , \quad
 \epsilon _{\text{bgr}}=1.
\end{eqnarray}
The solution of (\ref{sigma eom}) for $\sigma$ is  given in terms of Bessel functions with the asymptotic limit, 
\begin{equation}
\label{sigma curv}
\sigma \approx \sigma _0\left(\frac{t}{t_0}\right)^{-3/2}\left(\sin \left(m_{\sigma }t+\alpha \right)+\frac{3}{8m_{\sigma }t}\cos \left(m_{\sigma
}t+\alpha \right)\right),
\end{equation}
for $m_{\sigma }t\gg 1$ and $\alpha$ a phase set be the details of the initial excitation mechanism; the latter is of little interest to us (we often set $\alpha=0$ for simplicity).  The subscript ``$_0$'' indicates the moment of the initial excitation, e.g.~the tunnelling event. The maximal oscillation amplitude is constrained by (\ref{constraintsigma0a}) and (\ref{constraintsigma0b}) (for larger amplitudes, multi-field effects become important at the perturbed level, which can not be recovered in our approach). The resulting oscillatory contribution to the Hubble parameter becomes
\begin{equation}
\label{Hosci curv}
H_{\text{osci}}=-\frac{\sigma _0^2m_{\sigma }}{8 }
\left(\frac{t}{t_0}\right)^{-3}\sin \left(2 \left(m_{\sigma }t+\alpha \right)\right)\,,
\end{equation}
and the leading oscillatory components of the slow-roll parameters read
\begin{equation}
\label{eps osci curv}
\epsilon _{\text{osci}}=\frac{\sigma _0^2m_{\sigma }^2}{4 H_{\text{bgr}}^2 }
\left(\frac{t}{t_0}\right)^{-3}\cos \left(2
\left(m_{\sigma }t+\alpha \right)\right),
\end{equation}
\begin{equation}
\label{eta osci curv}
\left(\epsilon \dot{\eta }\right)_{\text{osci}}=-\frac{\sigma _0^2m_{\sigma }^4}{H_{\text{bgr}}^3
}
\left(\frac{t}{t_0}\right)^{-3}\cos
\left(2 \left(m_{\sigma }t+\alpha \right)\right).
\end{equation}
Using the above in the power-spectrum correction (\ref{power simple}) and the shape function (\ref{shape simple}) leads to
\begin{eqnarray}
 \frac{\Delta P_\zeta}{P_{\zeta}}(k)&=&\frac{\sigma _0^2m_{\sigma }^2}{4 H_{\text{bgr}}^2\left(t_0\right)
}
\int _{\tau_\text{min} }^{\tau_\text{max}}
\left(\frac{t}{t_0}\right)^{-1}\\&&
\nonumber\cos \left(2 \left(m_{\sigma }t+\alpha \right)\right)\left(\frac{1-2 k^2}{k}\sin (2k\tau )+ 2\cos (2 k \tau)\right)d\tau ,\\
S\left(k_1,k_2,k_3\right)&=&
k_ 1 k_2 k_3 \mathcal{N}\,
\frac{\sigma_ 0^2 m_\sigma ^4}
{16  H_{\text{bgr}}^4(t_0)}\\&&\nonumber 
\int _{\tau_\text{min} }^{\tau_\text{max}}
 (2K \cos (K \tau)-6 \sin (K \tau))\cos(2(m_\sigma t +\alpha))d\tau,
\end{eqnarray}
where we recall that all $k_i$ are rescaled with $k_\text{eq}$, we used $K=(k_1+k_2+k_3)$ and defined
\begin{eqnarray}
\mathcal{N}(t_\text{eq},t_0)\equiv \left(\frac{H_\text{bgr}(t_\text{eq})a(t_\text{eq})}{H_\text{bgr}(t_0)a(t_0)}\right)^3\,.\label{def:mathcalN}
\end{eqnarray}
Since $aH_\text{bgr}=1$ during the curvature dominated regime we have $\mathcal{N}=1$. 
 After substituting the integration boundaries of the curvature dominated regime, 
\begin{eqnarray} 
 \tau_\text{min}=\tau_i^{(k)}\quad ,\quad \tau_\text{max}=\tau_f^{(k)}
 \end{eqnarray} 
 from Sec.~\ref{sec:history}, and applying the stationary phase approximation,  
 the power-spectrum becomes for high frequencies of the arguments of the trigonometric functions (large $k$ and/or $m_\sigma$) 
\begin{eqnarray}
\label{power curv}
\nonumber \frac{\Delta P_\zeta}{P_{\zeta}}(k)&\approx& 
\frac{\sqrt{\pi }}{4}
\sigma_0^2\frac{m_\sigma^3}{H_{\text{bgr}}^3(t_0)}
k^{-3/2}
\bigg( \frac{1-2k^2}{2k}\sin \left(2k\left(1- \omega(k)\right)+2\alpha +\frac{\pi }{4}\right) \\&&
+\cos\left(2k\left(1- \omega(k)\right)+2\alpha +\frac{\pi }{4}\right)\bigg)\,.
\end{eqnarray}
where the frequency is set by the 
\begin{eqnarray}
\omega(k)\equiv \ln\left(\frac{k\sqrt{\rho_\phi/3}} {m_\sigma}\right)\,.\label{def:omega}
\end{eqnarray}
\begin{figure}[t]
        \begin{subfigure}[b]{0.5\textwidth}
                \centering
                \setlength\fboxsep{0pt}
                \setlength\fboxrule{0.5pt}
                \includegraphics[width=\textwidth]{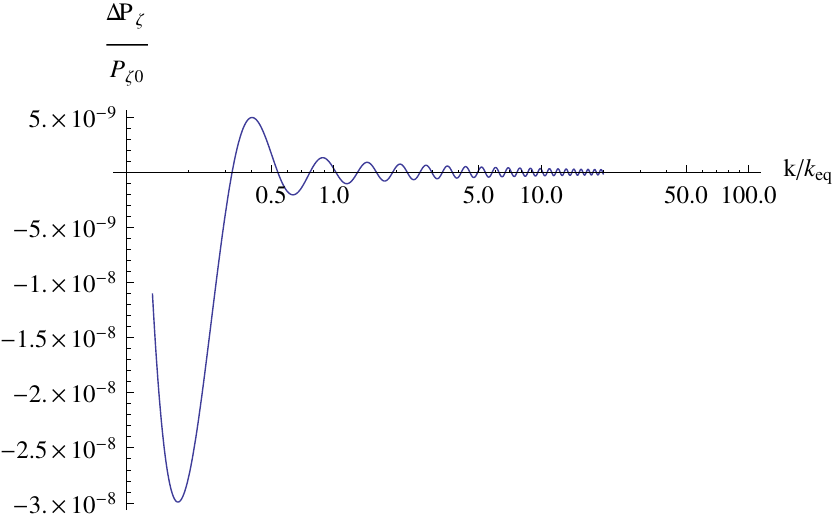}
                \caption{
                 $m_\sigma=200\sqrt{\frac{\rho_\phi}{3}}$, $\sigma_0=0.001 $, $t_\text{i}=\frac{1}{1500}\sqrt{\frac{3}{\rho_\phi}}$
                }
                \label{fig:power big time curv}
        \end{subfigure}
                ~ 
                \begin{subfigure}[b]{0.5\textwidth}
                        \centering
                        \setlength\fboxsep{0pt}
                        \setlength\fboxrule{0.5pt}
                        \includegraphics[width=\textwidth]{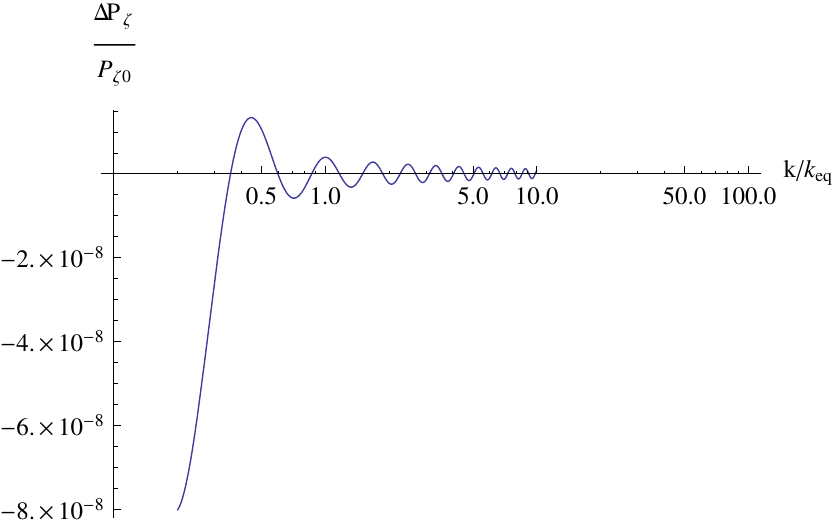}
                        \caption{\small$m_\sigma=100\sqrt{\frac{\rho_\phi}{3}}$,   $\sigma_0=0.001 $, $t_\text{i}=\frac{1}{500}\sqrt{\frac{3}{\rho_\phi}}$}
                        \label{fig:power eq time curv}
                \end{subfigure}
                ~ 

                  		\centering
        \begin{subfigure}[b]{0.5\textwidth}
                \centering
                \setlength\fboxsep{0pt}
                \setlength\fboxrule{0.5pt}
                \includegraphics[width=\textwidth]{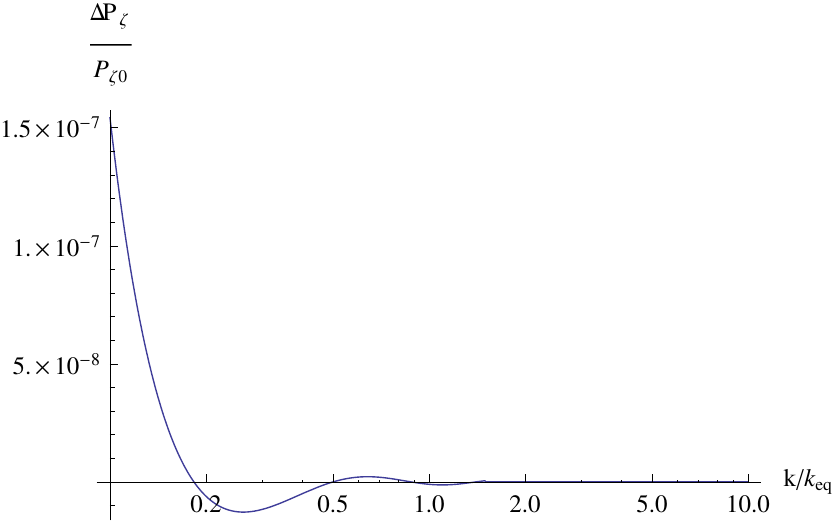}
                \caption{\small{$m_\sigma=15 \sqrt{\frac{\rho_\phi}{3}}$, $\sigma_0=0.01 $},   $t_\text{i}=\frac{1}{500}\sqrt{\frac{3}{\rho_\phi}}$}
                \label{fig:power small time curv}
        \end{subfigure}
        \caption{\small{The power-spectrum in the curvature dominated regime for different masses of the $\sigma$-field plotted over the physical wave-number $k$, normalised by the horizon scale at the time of equality $k_\text{eq}=a(t_\text{eq})H_\text{bgr}(t_\text{eq})$, with $\alpha=0$. The time of excitation $t_\text{i}$ is taken to be after the tunnelling event. 
        }} 
        \label{fig:power curv}
\end{figure}
Similarly, the shape function can be approximated by
\begin{eqnarray}
\nonumber S\left(k_1,k_2,k_3\right)&\approx&
k_ 1 k_2 k_3\mathcal{N}\,
\frac{\sigma_ 0^2}{8 }\frac{ m_\sigma^4}
{H_{\text{bgr}}^4(t_0)}\sqrt{\frac{\pi K}{2}}
\bigg(\cos \left( K\left(1- \omega(K/2))\right)+2\alpha +\frac{\pi}{4}\right)\\&&
 +\frac{3}{K} \sin \left( K\left(1- \omega(K/2))\right)+2\alpha +\frac{\pi}{4}\right)\bigg).
 \label{shape curv}
\end{eqnarray}

\begin{figure}[h]  
      \begin{subfigure}[b]{0.5\textwidth}
                \centering
                \setlength\fboxsep{0pt}
                \setlength\fboxrule{0.5pt}
                \includegraphics[width=\textwidth]{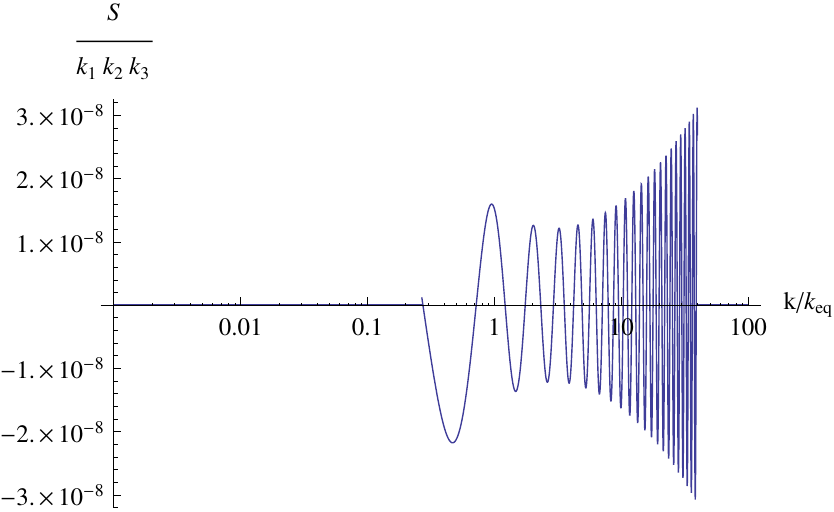}
                \caption{\small $m_\sigma=200\sqrt{\rho_\phi/3}$, $\sigma_0=0.01 $}
                \label{fig:bi big time curv}
        \end{subfigure}
        ~
        ~ 
        \begin{subfigure}[b]{0.5\textwidth}
                \centering
                \setlength\fboxsep{0pt}
                \setlength\fboxrule{0.5pt}
                \includegraphics[width=\textwidth]{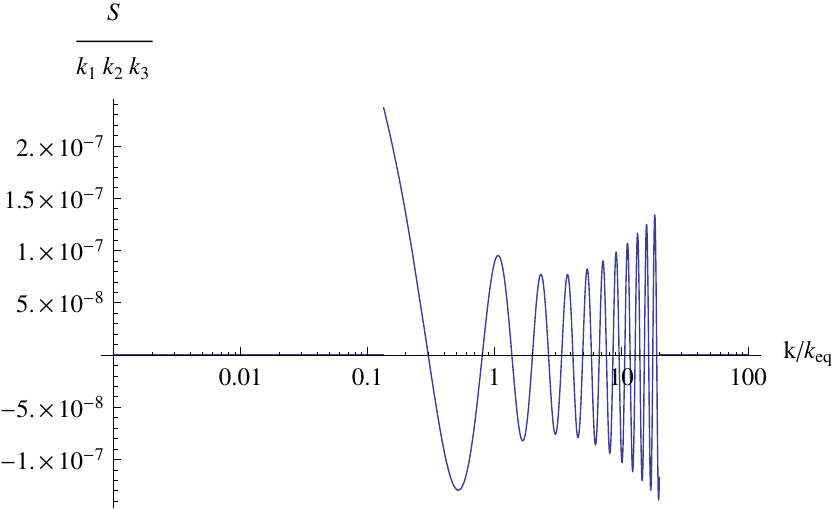}
                \caption{\small $m_\sigma=100\sqrt{\rho_\phi/3}$, $\sigma_0=0.1 $}
                \label{fig:bi eq time curv}
        \end{subfigure}
        ~ 

\centering
        \begin{subfigure}[b]{0.5\textwidth}
                \centering
                \setlength\fboxsep{0pt}
                \setlength\fboxrule{0.5pt}
                \includegraphics[width=\textwidth]{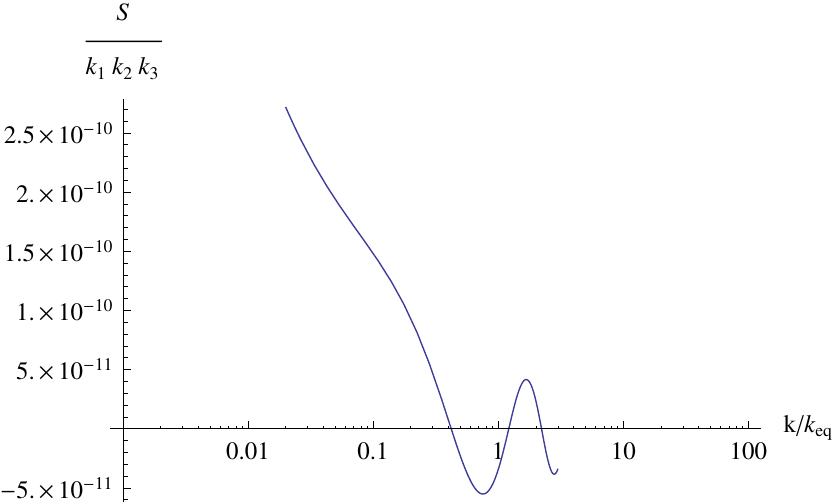}
                \caption{\small $m_\sigma=15\sqrt{\rho_\phi/3}$, $\sigma_0=0.1 $}
                \label{fig:bi small time curv}
        \end{subfigure}
        \caption{\small The shape function in the curvature dominated regime for different masses of the $\sigma$-field plotted over the physical wave-number $k$, normalised by the horizon scale at the time of equality $k_\text{eq}$, with $\alpha=0$. The time of excitation $t_\text{i}=(1500\sqrt{\rho_\phi/3})^{-1}$ is taken to be after the tunnelling event but well before the end of the curvature dominated regime. 
        }
        \label{fig:bispec curv}
\end{figure}

We plot the resulting oscillatory corrections to $P_\zeta$ in Fig.~\ref{fig:power curv} for exemplary choices of the excitation time $t_\text{i}$ (we allow $t_\text{i}$ to differ from $t_0$, in which case the excitation is not a result of the tunnelling event), initial amplitude $\sigma_0$ and mass $m_\sigma$, making sure that $\sigma_0$ satisfies (\ref{constraintsigma0a}) and (\ref{constraintsigma0b}).
Evidently, the oscillation amplitude is proportional to $\sigma_0^2$ with a decaying envelope at large wave numbers, caused by the $f$-terms in (\ref{power simple}), with $f(\tau)$ defined in (\ref{f}). Corresponding oscillations are present in $S$, plotted in Fig.~\ref{fig:bispec curv}. However, its envelope has a minimum due to the decaying sine mode induced by the $f(\tau)$ term. For comparison, the corresponding cosine mode in the power-spectrum has little effect, because it is sub-dominant over the entire k-domain.

For both, power-spectrum and bi-spectrum, the cut-off in the signal is determined by the length of the curvature dominated period between
\begin{eqnarray}
1=a^{(k)}_\text{0}\leq a_\text{i}\leq a \leq a^{(k)}_\text{f}=\frac{1}{A\sqrt{\rho_\phi/3}}\,.
 \end{eqnarray} 
Since the resonating wave-number leading to the oscillations is given by $k=a m_\sigma$, the boundaries of the scale factor translate to the interval 
\begin{eqnarray}
m_\sigma\leq k \leq \frac{m_\sigma}{A\sqrt{\rho_\phi/3}}\,,\label{intervalk}
\end{eqnarray}
 for our rescaled wave-number  if $a_\text{i}= a^{(k)}_0$ (and a shorter interval if the oscillations start at $t_\text{i}>t_0$), where $k$ stands for $K/2$ in the shape function case. 
 
 It is instructive to compare the period of the $\sigma$-field oscillations $T=2\pi/ m_\sigma$ with the available time $t_\text{f}-t_0= (A^2\rho_\phi/3)^{-1/2} -1$. If a $\sigma$-oscillation can't be completed before the end of the regime, a resonance with the vacuum oscillations is impossible and no power-spectrum signal is generated. However, since the individual resonating modes $k_i$ involved in the bi-spectrum calculation are necessarily smaller than the corresponding power-spectrum modes, they have a chance to resonate with the $\sigma$-field, leading to a signal in the shape function, even if none is present in the power-spectrum. Given a few oscillations, a small resonant running arises.

\subsubsection{A Case Study and Observational Prospects \label{sec:casestudy}}
It is instructive to consider a concrete inflationary setup, particularly to estimate the maximally attainable signals in the power- and bi-spectrum. 

As in Sec.~\ref{sec:truncating}, we use a simple  inflationary model with $V(\phi)=m_\phi^2\phi^2/2$, $m_{\phi}\sim \sqrt{8\pi}\times 10^{-6}$ and $\phi\approx 16 $ in the slow-roll limit, so that around sixty e-folds of inflation result after curvature becomes sub-dominant\footnote{That is, we put the observational window of the CMB around $k_\text{eq}$. Since signals are generated deep inside the horizon during curvature domination, relevant modes can cross the horizon during inflation and lie in the observational window.} and the normalization of the power-spectrum is satisfied. The curvature dominated regime ends at
\begin{eqnarray}
a_\text{f}^{(k)}\approx \frac{\sqrt{6}}{Am_\phi\phi}\approx 3.0\times 10^3 \,,
\end{eqnarray}
where we used $A=10$, which implies an interval in (\ref{intervalk}) of
\begin{eqnarray}
m_\sigma\leq k\lesssim \frac{m_\sigma}{m_\phi}0.015\label{intervallcasestudy}
\end{eqnarray}
where a signal could be generated, and an oscillation frequency in (\ref{def:omega}) of
\begin{eqnarray}
\omega(k)=\ln\left(k\frac{m_\phi}{m_\sigma\sqrt{6}}\phi\right)\approx \ln\left(6.5\, k \frac{m_\phi}{m_\sigma}\right)\,.\label{defomegacase1}
\end{eqnarray}

For which $m_\sigma$ can a signal be generated in the power-spectrum? Allowing for at least one full oscillation of $\sigma$ in the curvature dominated regime leads to the condition $2\pi/m_\sigma \leq t_\text{f}-t_\text{0}=a_\text{f}^{(k)}-a_\text{0}^{(k)}$, which becomes
\begin{eqnarray}
\frac{m_\sigma}{m_\phi}&\geq & \frac{2\pi A\phi}{\sqrt{6}-Am_\phi\phi}\approx 2\pi A\frac{\phi}{\sqrt{6}}
\approx  410\,.
\end{eqnarray}
Thus, the tunnelling fields mass needs to be bigger than 
\begin{eqnarray}
m_\sigma^\text{min}\equiv \frac{2\pi A\phi}{\sqrt{6}-Am_\phi\phi}m_\phi\,,
\end{eqnarray}
considerably heavier than the mass of the inflaton. This is not surprising, since otherwise $\sigma$ is frozen during the curvature dominated regime due to Hubble friction, just like the inflaton.

Assuming $m_\sigma>m_\sigma^\text{min}$, how big of a signal can be generated?  Since $\delta\leq 0.1$, see eqn.~(\ref{defdelta}), and $m_\phi \ll m_\sigma$, the strongest constraint on the oscillation amplitude of $\sigma$ stems from (\ref{constraintsigma0a}), that is
\begin{eqnarray}
\sigma\ll\frac{m_\phi}{m_\sigma}\frac{1}{\delta}\,,
\end{eqnarray}
which needs to be satisfied throughout the oscillations of $\sigma$. Violating this bound does not mean that signals are absent, but that they can not be computed in the truncated single-field description at the perturbed level, which we employ.

Since $\delta(t) \propto a(t)$ and the oscillation amplitude of $\sigma$ redshifts as $a^{-3/2}$, the bound remains satisfied, if it is satisfied initially.
 Hence, we need to impose
\begin{eqnarray}
\sigma_0\ll\frac{m_\phi}{m_\sigma}\frac{1}{\delta_\text{i}}\,,
\end{eqnarray}
at the time of the initial excitation. The farther the excitation lies in the past, the smaller is the observable signal. 
To provide a conservative upper bound on amplitudes, we consider an initial excitation of $\sigma$ at exactly one oscillation period before the end of the curvature dominated regime,
\begin{eqnarray}
T=\frac{2\pi}{m_\sigma}=t_\text{f}-t_\text{i}=a_\text{f}^{(k)}-a_\text{i}\,,
\end{eqnarray}
 which determines $a_\text{i}$; thus, we arrive at
\begin{eqnarray}
\delta_\text{i}=\frac{a_\text{i}}{a_\text{f}^{(k)}}\delta_\text{f}^{(k)}\,,
\end{eqnarray}
with $\delta_\text{f}^{(k)}=1/\sqrt{A^2+1}\approx 0.1$. Such a single period in $\sigma$ generates a  signal close to the upper boundary of the interval in (\ref{intervallcasestudy}) only (i.e.~in Fig.~\ref{fig:power curv} before the upper cut-off on $k$).

Using $3H_\text{bgr}^2(t_0)\approx V(\phi)/\delta_\text{i}^2\gg V(\phi)$ and ignoring factors of order one, we can put an upper bound on the amplitude of the power-spectrum correction in (\ref{power curv})  
\begin{eqnarray}
\frac{\Delta P_\zeta}{P_\zeta}&\ll& \delta_\text{i}\frac{m_\sigma}{m_\phi}\frac{1}{\phi^3}\\
&=&\frac{m_\sigma^\text{min}}{m_\phi}\frac{1}{\phi^3\sqrt{A^2+1}}\left(\frac{m_\sigma}{m_\sigma^\text{min}}+\left(\frac{Am_\phi\phi}{\sqrt{6}}-1\right)\right)\\
&\approx& 0.01\left(\frac{m_\sigma}{m_\sigma^\text{min}} -0.9997\right)\,, \label{limitpowerecaseA}
\end{eqnarray}
and the bi-spectrum in (\ref{shape curv}) 
\begin{eqnarray}
\frac{S}{k_1k_2k_3}&\ll& \mathcal{N}\delta_\text{i}^2\frac{m_\sigma^2}{m_\phi^2}\frac{1}{\phi^4}=\mathcal{N}\left(\frac{\Delta P_{\zeta}}{P_{\zeta}}\right)^2\phi^2\,.
\end{eqnarray}
where $\mathcal{N}=1$. Thus, a correction at the percent level in the power-spectrum is possible if the initial excitation happens close to $t_\text{f}$ and $m_\sigma$ is large enough; of course, any mass should be smaller than the Planck mass,
\begin{eqnarray}
m_\sigma\ll 1\,,
\end{eqnarray}
yielding $\Delta P_\zeta/P_\zeta \ll 5$ in our toy model.
 In the plots of Fig.~\ref{fig:power curv}, we chose earlier times  to cover several oscillations, and masses well below the Planck mass, resulting in weak signals. $\sigma_0$ satisfies the bound in (\ref{constraintsigma0a}) in all plots; in these cases the resulting amplitudes are much smaller than the above bound. However, even if the correction to the power-spectrum is about $10\%$ (possible to generate with $m_\sigma$ close to the Planck mass and tuning of $t_\text{i}$, while still marginally allowed by current data), the bi-spectrum remains dauntingly small, $S/(k_1k_2k_3)\sim \mathcal{O}(1)$.
Further, as $m_\sigma$ is increased,  the frequency $\omega$ in (\ref{def:omega}) grows too, complicating a detection in a noisy data set.

Let us estimate the amplitude of the power-spectrum correction if the initial excitation is due to a displacement of $\sigma$ directly after the tunnelling event. Using the same line of thought as above with $a_\text{i}=a_0=1$ leads to
\begin{eqnarray}
\frac{\Delta P_\zeta}{P_\zeta}&\ll& \frac{\delta_\text{f}^{(k)}}{a_\text{f}^{(k)}}\frac{m_\sigma}{m_\phi}\frac{1}{\phi^3}\\
&\approx& 3.3 \times 10^{-6}\frac{m_\sigma}{m_\sigma^{\text{min}}}\,.
\end{eqnarray}
Unfortunately, such a signal is currently unobservable: even if $m_\sigma=1$, we have $\Delta P_\zeta/P_\zeta \ll 0.0016$.

Lastly, let us have a brief look at current observations. The amplitude of any oscillations on top of the power-spectrum of the WMAP satellite has to be lower than about 10\% \cite{Meerburg:2011gd,Chen:2012ja,Saito:2012pd,Benetti:2012wu}, depending on the frequency. Such oscillations in the power-spectrum are conceivable, but do not improve the theoretical fit significantly \cite{Meerburg:2011gd,Benetti:2012wu}. Using the flat sky approximation
$d_A=l/k$, where $d_A=13.7 \text{ Gpc}$, and the pivotal scale $k_\text{pivot}=0.027 \text{ Mpc}^{-1}$, we can estimate the maximal frequency resolution to 
\begin{eqnarray}
|\omega| \ll \frac{1162}{s}\,,
\end{eqnarray}
where
 \begin{eqnarray}
 s\equiv \frac{k_\text{pivot}}{k_\text{eq}}\,,
 \end{eqnarray}
and we required that an oscillation stretches over at least two $l$-modes in order to be observable. Realistically, some binning of the data is necessary to reduce the noise. Combining 10 $l$-modes into a single bin, the maximal frequency is reduced by a factor of $10$; hence, we define 
\begin{eqnarray}
\omega_\text{obs}^\text{max}\equiv \frac{116}{s} \,.
\end{eqnarray}
We can tune $s$, and thus the position of this window
\begin{eqnarray}
k_\text{obs} \in (0.0054, 3.2)\,s\,,
\end{eqnarray}
by moving the initial value of the inflaton and/or its potential after the tunnelling event, as long as the subsequent inflationary phase last long enough to dilute curvature below current observational limits. Naturally, we are primarily interested in $s\sim \mathcal{O}(1)$, since otherwise any trace of the curvature dominated regime is pushed out of the observable range.

Fortunately, the frequency in (\ref{defomegacase1}) is only logarithmically dependent on $k$ and $m_\phi/m_\sigma$: taking $s=1$, $k=0.0054$ and $m_\sigma=m_\sigma^\text{min}$ yields the estimate
\begin{eqnarray}
|\omega|\lesssim 9.4\,,
\end{eqnarray}
well below $\omega_\text{obs}^\text{max}$. 

To summarize, the main challenge to observe an oscillatory signal in the power-spectrum generated during the curvature dominated regime is the generically small amplitude: an observation would indicate an excitation shortly before $t_\text{eq}$ with a near maximally allowed amplitude and a large $m_\sigma$. However, the frequency of oscillations is commonly low enough to pose no additional challenge.  The generated bi-spectrum is proportional to $\Delta P_\zeta^2/P_\zeta^2$ and thus  unobservable with current experiments. 

\subsection{Inflation \label{sec:infl dom}}

The inflationary regime has been discussed in \cite{Chen:2011zf}, whose results we recover and extend slightly. Neglecting the contribution from curvature, we consider a simple power-law inflationary regime,
\begin{equation}
a \propto  t^p\,.
\end{equation}
 Conformal time is given by
\begin{equation}
\tau =\frac{1}{(1-p)}\frac{t}{a} \,.
\end{equation}
As $p\to 1$, a conformal singularity is approached, indicating the need to treat the transition region to curvature domination as a separate case. 

Using the Hubble parameter $H_{\text{bgr}}=p/t$, the solution of the $\sigma$-field's equation of motion is
\begin{equation}
\label{sigma infl}
\sigma \approx \sigma _0\left(\frac{t}{t_0}\right)^{-3 p/2}\left(\sin \left(m_{\sigma }t+\alpha \right)+\frac{9p^2-6p}{8m_{\sigma }t}\cos \left(m_{\sigma
}t+\alpha \right)\right)\,.
\end{equation}
The initial amplitude may be set during inflation or in a prior regime, in which case $t_0$ coincides with the beginning of the inflationary era.
We only focus on the latter case in this article, keeping in mind that a generalization to the former one is straightforward. Thus, the leading order oscillatory contribution to the Hubble parameter is
\begin{equation}
\label{H osci infl}
H_{\text{osci}}=-\frac{\sigma _0^2m_{\sigma }}{8 }\left(\frac{t}{t_0}\right)^{-3p}\sin \left(2 \left(m_{\sigma }t+\alpha \right)\right),
\end{equation}
so that the leading order oscillations of the slow-roll parameters become
\begin{eqnarray}
\label{eps osci infl}
\epsilon _{\text{osci}}(t)&=&\frac{\sigma _0^2m_{\sigma }^2}{4 H_{\text{bgr}}^2 }\left(\frac{t}{t_0}\right)^{-3p}\cos \left(2
\left(m_{\sigma }t+\alpha \right)\right)\,,
\label{eta osci infl}\\
\left(\epsilon\dot{\eta }\right)_{\text{osci}}(t)&=&-\frac{\sigma _0^2m_{\sigma }^4}{H_{\text{bgr}}^3 }\left(\frac{t}{t_0}\right)^{-3p}\cos
\left(2 \left(m_{\sigma }t+\alpha \right)\right)\,.
\end{eqnarray}
Inserting the above in the power-spectrum correction (\ref{power simple}) and the shape function (\ref{shape simple}) leads to
\begin{eqnarray}
\frac{\Delta P_\zeta}{P_{\zeta}}(k)&=&\frac{\sigma _0^2m_{\sigma }^2}{4 H_{\text{bgr}}^2\left(t_0\right)
\epsilon _{\text{bgr}}}
\int _{\tau_\text{min}}^{\tau _\text{max}}\left(\frac{t}{t_0}\right)^{-3p+2}
\cos \left(2 \left(m_{\sigma }t+\alpha \right)\right)\\&&\nonumber
\left(\left(\frac{p^2}{k (1-p)^2\tau^2}-2 k\right)\sin(2k \tau)+\frac{2p}{(1-p)\tau}\cos (2 k \tau)\right)d\tau\,,
\end{eqnarray}
and
\begin{eqnarray}
S\left(k_1,k_2,k_3\right)&=&
\mathcal{N}k_ 1 k_2 k_3\,
\frac{\sigma_ 0^2 m_\sigma^4}
{8 H_{\text{bgr}}^4(t_0)  \epsilon_\text{bgr}}
\int _{\tau_\text{min} }^{\tau_\text{max}}
\left(\frac{t}{t_0}\right)^{3-3p}\\&&\nonumber
\left(K \cos (K \tau)-\frac{3 p}{(1-p)\tau} \sin (K \tau)\right)\cos(2(m_\sigma t +\alpha))d\tau\,,
\end{eqnarray}
respectively, where we used the definition of the normalization constant $\mathcal{N}$ from (\ref{def:mathcalN}), reading
\begin{eqnarray}
\mathcal{N}(t_\text{eq},t_0)=\left(\frac{t_\text{eq}}{t_0}\right)^{3(p-1)}\,,
\end{eqnarray}
for power-law inflation. Based on the stationary phase approximation, the asymptotic behaviour of the power-spectrum correction 
\begin{eqnarray}
\label{power infl asym}
\frac{\Delta P_\zeta}{P_\zeta}(k)&\approx & \frac{\sqrt{\pi}}{4} \frac{1}{\epsilon_\text{bgr}}\sigma_0^2
\left(\frac{m_\sigma}{H_{\text{bgr}}\left(t_0\right)}\right)^{\frac{5}{2}}
\left(\frac{2 k}{k_r}\right)^{-3+\frac{5}{2p}}\\&& \nonumber
\left(\left(1-\frac{H_\text{bgr}^2(t_0)}{2 m_\sigma^2}\left(\frac{2 k}{k_r}\right)^{-\frac{2}{p}}\right)\sin\left(\omega(k)\frac{2k}{k_r}+2\alpha+\frac{\pi }{4}\right)\right.\\
 && \nonumber+\left.\frac{1}{k}
\left(\frac{2 k}{k_r}\right)^{-\frac{1}{p}} \frac{H_\text{bgr}(t_0)}{m_\sigma}\cos\left(\omega(k)\frac{2k}{k_r}+2\alpha+\frac{\pi }{4}\right)\right)\,,
\end{eqnarray}
and the shape function
\begin{eqnarray}
\label{shape infl asym}
&&S(k_1, k_2, k_3)\approx k_1k_2k_3\mathcal{N}
\frac{\sqrt{\pi }}{8}
\frac{1}{\epsilon_\text{bgr}}
\sigma_0^2
\left(\frac{m_\sigma}{H_\text{bgr}(t_0)}\right)^{\frac{9}{2}}
\left(\frac{K}{k_r}\right)^{-3+\frac{7}{2p}} \\
\nonumber && \quad \quad \quad \quad \times \left(\cos\left(\omega\left(K/2\right)\frac{K}{k_r}+ 2\alpha+\frac{\pi }{4}\right)+ \frac{3 H_\text{bgr}}{2 m_\sigma}\left(\frac{K}{k_r}\right)^{-\frac{1}{p}}\sin \left(
\omega\left(K/2\right)\frac{K}{k_r}+ 2\alpha +\frac{\pi }{4}\right)\right)\,,
\end{eqnarray}
is again oscillatory, as in the curvature dominated phase, albeit with a different frequency 
 \begin{eqnarray}
 \omega(k)\equiv p\left| \frac{p}{1-p}\right| \frac{2m_{\sigma }}{H_\text{bgr}(t_0)}\left(\frac{2k}{k_r}\right)^{1/p-1}\,,
 \end{eqnarray}
and pre-factor. Following \cite{Chen:2011tu}, we denoted the first mode that starts to resonate as soon as the $\sigma $-field is excited by
\begin{eqnarray}
 k_r\equiv 2 m_\sigma a_0\,.
\end{eqnarray}
  
We note two minor differences as compared to \cite{Chen:2011tu}: firstly,  we retained the $f$-term contributions to the power-spectrum corrections, providing the two sub-dominant terms in (\ref{power infl asym}). Secondly, just as in the curvature dominated regime, 
we retain a dependence on $k_1 k_2 k_3$  of $S$ stemming from the first term in (\ref{shape infl asym}) tracing back to our chosen definition of the shape function. 
Retaining the dominant term only, we fully recover Chen's results. 

\begin{figure}[t]  
      \begin{subfigure}[b]{0.5\textwidth}
                \centering
                \setlength\fboxsep{0pt}
                \setlength\fboxrule{0.5pt}
                \includegraphics[width=\textwidth]{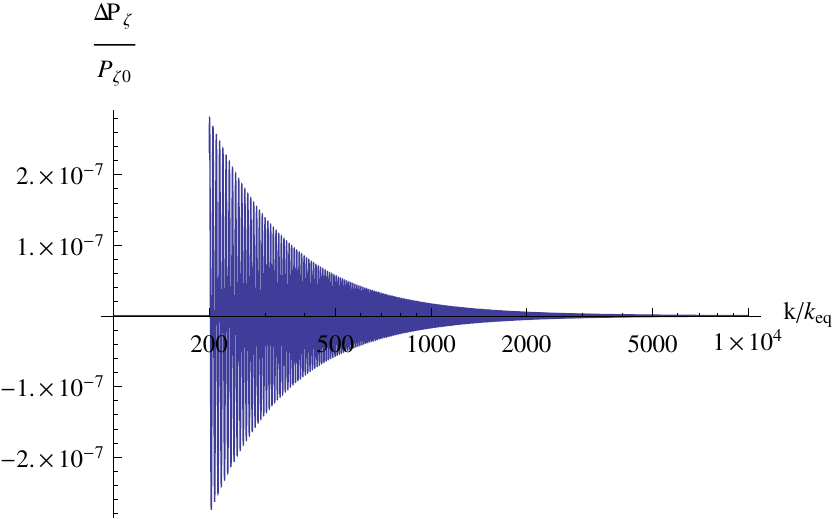}
                \caption{\small$m_\sigma/\sqrt{\rho_\phi/3}=20$, $p=2$, $\sigma_0=0.0001 $}
                \label{fig:power big time infl}
        \end{subfigure}
        ~
        ~ 
        \begin{subfigure}[b]{0.5\textwidth}
                \centering
                \setlength\fboxsep{0pt}
                \setlength\fboxrule{0.5pt}
                \includegraphics[width=\textwidth]{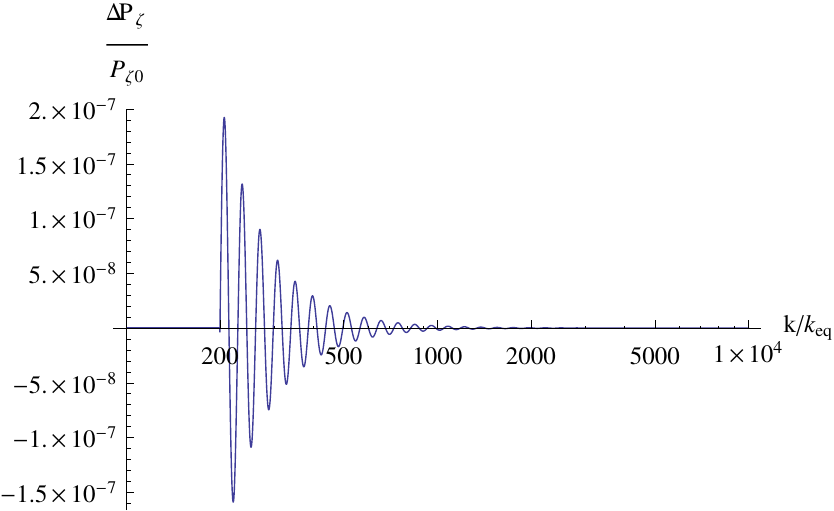}
                \caption{\small  $m_\sigma/\sqrt{\rho_\phi/3}=20$, $p=20$, $\sigma_0=0.1 $}
                \label{fig:power eq time infl}
        \end{subfigure}
        ~ 

\centering
        \begin{subfigure}[b]{0.5\textwidth}
                \centering
                \setlength\fboxsep{0pt}
                \setlength\fboxrule{0.5pt}
                \includegraphics[width=\textwidth]{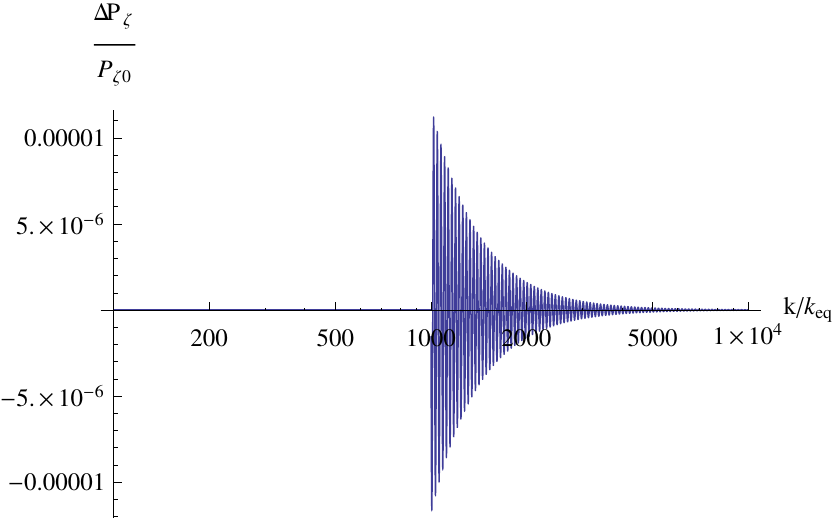}
                \caption{\small $m_\sigma/\sqrt{\rho_\phi/3}=100$, $p=20$, $\sigma_0=0.01 $}
                \label{fig:power small time infl}
        \end{subfigure}
        \caption{\small Power-spectrum corrections during the inflationary regime with $\epsilon_\text{bgr}=p^{-1}$, $t_0=t_\text{eq}$, $\alpha=0$, and $k_r=2 m_\sigma a_0$ for different values of $m_\sigma/\sqrt{\rho_\phi/3}$ and $p$.}\label{fig:power infl}
\end{figure}

The expressions (\ref{power infl asym}) and (\ref{shape infl asym}) are valid during inflation
\begin{eqnarray}
a_\text{i}^\phi\leq a\leq a_\text{f}^\phi 
\end{eqnarray}
with $a_\text{i}^{\phi}$ from (\ref{aiphi}) as well as $B=0.1$, and
\begin{eqnarray}
 a_\text{f}^\phi\equiv a_\text{i}^\phi e^{N+\ln(B)}\gtrsim a_\text{i}^\phi e^{60}\,,
\end{eqnarray}
 which translates to the interval 
\begin{eqnarray} 
 \frac{m_\sigma}{B\sqrt{\rho_\phi/3}}<k<\frac{m_\sigma}{B\sqrt{\rho_\phi/3}}e^{60}\,,
 \end{eqnarray}
 where a signal could be generated. Here $k$ stands again for $K/2$ in the shape function case.
The boundary is only important at the low end of the spectrum, since the exponential factor sends the upper boundary to unobservably high wave numbers. Physically, the $\sigma$-oscillations resonate only with modes that haven't crossed the horizon yet, so that we only need to be careful about the lower bound of the interval in contrast to the curvature dominated regime. If the excitation takes place late during inflation, the signal is shifted out of the observational window.
For a detection, the amplitude of the oscillations is of prime interest. We saw that the amplitude of oscillations caused by an initial displacement of $\sigma$ after the tunnelling event is too small at the end of the curvature dominated regime to be detectable in current experiments (within our computational framework). Hence, if an oscillation is detected, it must have its origin in a subsequent excitation event.
Further, the scale dependence of the signal's amplitude is  model dependent \cite{Chen:2011tu}: changing the slope or the width of the slow-roll valley can give qualitatively different envelopes of the oscillatory signal. 

\begin{figure}[t]
      \begin{subfigure}[b]{0.5\textwidth}
                \centering
                \setlength\fboxsep{0pt}
                \setlength\fboxrule{0.5pt}
                \includegraphics[width=\textwidth]{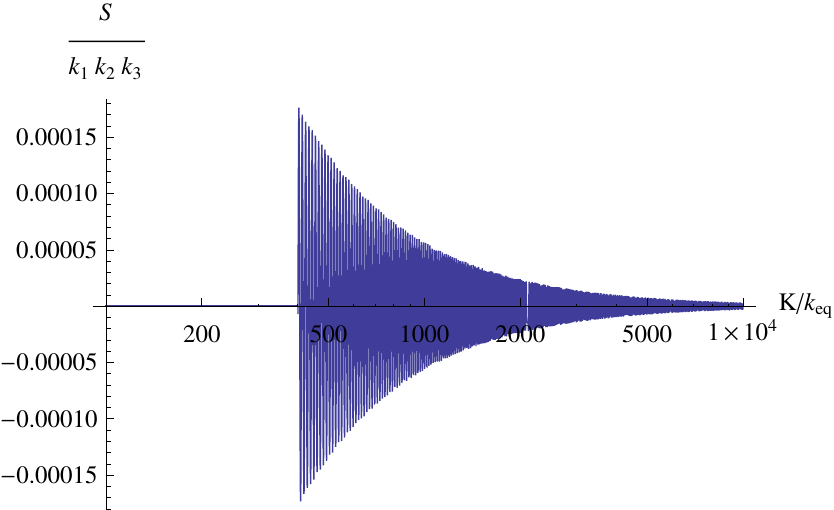}
                \caption{\small $m_\sigma/\sqrt{\rho_\phi/3}=20$, $p=2$}
                \label{fig:bi big time infl}
        \end{subfigure}
        ~
        ~ 
        \begin{subfigure}[b]{0.5\textwidth}
                \centering
                \setlength\fboxsep{0pt}
                \setlength\fboxrule{0.5pt}
                \includegraphics[width=\textwidth]{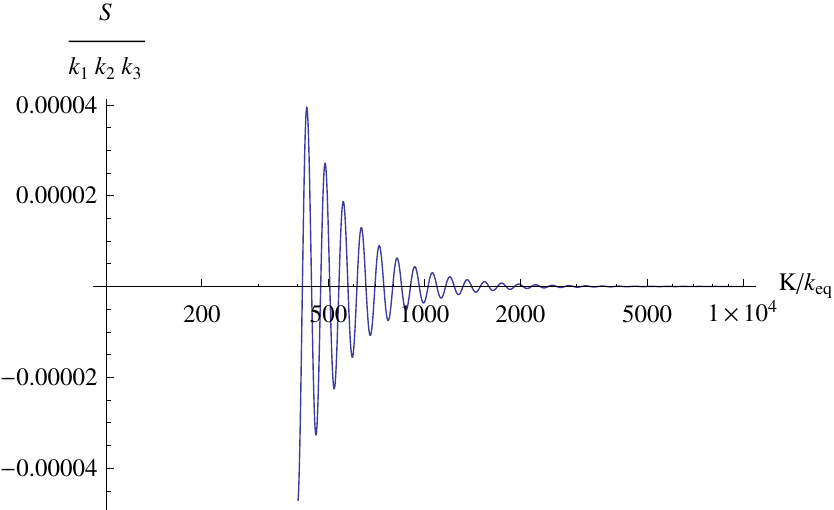}
                \caption{\small $m_\sigma/\sqrt{\rho_\phi/3}=20$, $p=20$}
                \label{fig:bi eq time infl}
        \end{subfigure}
        ~ 

          \centering
        \begin{subfigure}[b]{0.5\textwidth}
                \centering
                \setlength\fboxsep{0pt}
                \setlength\fboxrule{0.5pt}
                \includegraphics[width=\textwidth]{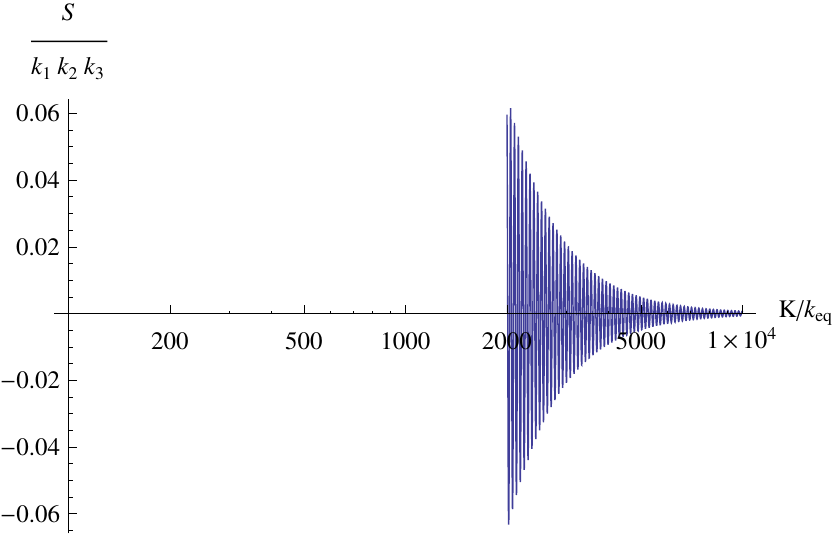}
                \caption{\small $m_\sigma/\sqrt{\rho_\phi/3}=100$, $p=20$}
                \label{fig:bi small time infl}
        \end{subfigure}
        \caption{\small Shape function during the inflationary regime with, $\epsilon_\text{bgr}=p^{-1}$, $t_0=t_\text{eq}$, $\sigma_0=0.0001 $, $\alpha=0$, and $k_r=2 m_\sigma a_0$ for different values of $m_\sigma/\sqrt{\rho_\phi/3}$ and $p$.}\label{fig:bispec infl}
\end{figure}

For the quadratic potential of $\sigma$, the amplitude of the bi-spectrum envelope decreases during inflation (Fig.~\ref{fig:bispec infl}), whereas it increases during the curvature dominated regime (Fig.~\ref{fig:bispec curv}) for all sets of parameters. Therefore, we expect to find a local maximum during the intermediate regime, which could act as a tell-tale sign of open inflation, if it were observed. 

The frequency $\omega(k)$ of the signal is again set by the inverse of the scale factor.  
Thus, if the amplitude is observable, the frequency of the oscillations acts as a direct probe of the scale factor's time dependence, as can be seen in Fig.~\ref{fig:power infl} and \ref{fig:bispec infl}. We refer the interested reader to \cite{Chen:2011zf,Meerburg:2011gd,Chen:2012ja,Saito:2012pd,Benetti:2012wu}, where signals generated during inflation and observational prospects are discussed in more detail.

Before investigating the intermediate regime, we would like to comment briefly on the allowed amplitudes for $\sigma_0$ during inflation. Since $\delta\approx 1$, the constraints on $\sigma_0$ in (\ref{constraintsigma0a}) and (\ref{constraintsigma0b}) read
\begin{eqnarray}
\sigma_0&\ll& \frac{m_\phi}{m_\sigma}\,,\label{constraintsigma0a2}\\
\sigma_0&\ll& \frac{m_\phi^2}{m_\sigma^2}\phi\,,\label{constraintsigma0b2}
\end{eqnarray}
where we used again $V(\phi)=m_\phi^2\phi^2/2$ as a prototypical inflationary potential. Some of the concrete case studies in Chen's work come dangerously close to these limits, particularly if $m_\sigma$ is large. For example, a case study in Sec.~6.3 of \cite{Chen:2011zf} (see also \cite{Chen:2012ja}) yielding a $10\%$ correction to the power-spectrum requires (leaving out all dimensionless factors of order one) $\beta\sim \dot{\phi}^2/(m_\sigma^2\sigma_0^2)\sim 10^{-2}$ 
during slow-roll, $\epsilon \sim \dot{\phi}^2/H^2\sim 10^{-2}$, which corresponds to an initial oscillation amplitude of $\sigma_0\sim 10^{-2}H /m_\sigma \sim 10^{-2}\phi m_\phi/m_\sigma$, 
where we used again our toy model with $\phi\approx 16$ in the last step to enable a comparison with our limits. While the first bound in  (\ref{constraintsigma0a2}) is marginally satisfied (one has to keep in mind that all dimensionless factors of order unity are ignored, which could easily lead to a factor of $10$), the second bound in (\ref{constraintsigma0b2}) becomes stronger for $m_\sigma/m_\phi>\phi\approx 16$ and is therefore violated easily for heavier masses. 

We conclude that if a detection of oscillations at the few percent level in the power-spectrum were to be made, and one wishes to attribute this signal to the oscillations of a heavy field during inflation, one should go beyond the current computational scheme and include perturbations in $\sigma$ throughout. 

\subsection{Intermediate Regime: Comparable Contributions \label{sec:com contr}}

In the intermediate regime, the contributions to the energy density from spatial curvature and the inflaton field are comparable. Since this interval is brief, Hubble friction is high, and $\sigma$ evolves much faster than $\phi$ since $m_\sigma\gg m_\phi$, we model the inflaton contribution as a constant in this section,
\begin{eqnarray}
\rho_\phi\approx \text{const.}\,,
\end{eqnarray}
 so that the solution of the Friedmann equation (\ref{Friedman}) is
\begin{equation}
\label{scale intermediate}
a(t)=\frac{\sinh(\sqrt{\rho_\phi/3} t)}{\sqrt{\rho_\phi/3}}\,,
\end{equation}
and the expression for conformal time $\tau$ takes the form
\begin{equation}
\tau=\ln\left(\tanh\left(\frac{\sqrt{\rho_\phi/3} t}{2}\right)\right).
\end{equation}
Note that the units of time are normalised differently -- with respect to the moment of equality rather than tunnelling, 
\begin{eqnarray}
 \tau_\text{eq}\equiv -\text{arcsinh}(1)\,,
 \end{eqnarray}
in order to map $t\in(0,\infty) \mapsto \tau \in(-\infty,0)$. Using (\ref{scale intermediate}), the background Hubble parameter becomes
\begin{equation}
H_\text{bgr}=\sqrt{\rho_\phi/3}\coth(\sqrt{\rho_\phi/3}  t)\,.\label{Hbgrinterm}
\end{equation}
Given this background, the equation of motion for $\sigma$ yields no simple analytic solution. In order to proceed, we expand the Hubble parameter around the time of equality
\begin{eqnarray}
t_\text{eq}=\frac{1}{\sqrt{\rho_\phi/3}}\text{arcsinh}(1)\,,
\end{eqnarray}
so that
\begin{equation}
H_\text{bgr}=\sqrt{\frac{\rho_\phi}{3}}\left(\sqrt{2}+\text{arcsinh}(1)-\sqrt{\frac{\rho_\phi}{3}} t\right)+\mathcal{O}\left((t-t_\text{eq})^2\right).
\end{equation}
Retaining the linear term is sufficient for our purposes, considering that $H_\text{bgr}$ acts merely as a friction term and we perform an asymptotic expansion later on. 

As before, in order to solve the equation of motion for the $\sigma$-field, we need to specify initial conditions. Technically, they are prescribed by the CdL instanton solution of the tunnelling event, dynamically evolved to the neighbourhood of $t_\text{eq}$ with the solution of the equation of motion during the curvature dominated era given by eqn.(\ref{sigma curv}).  However, provided that the frequency of oscillations in the curvature dominated era is large enough (which is the case), we can always choose to begin our considerations at the peak of on oscillations at $t_0$, where $\sigma=\sigma_0$ and $\dot{\sigma}_0=0$. 
 Further, we learned that an initial excitation directly after the tunnelling event is too weak to be observed; hence, we envision a subsequent excitation in the curvature dominated regime and treat $\sigma_0$ as a free parameter, only subject to the constraints in (\ref{constraintsigma0a}) and (\ref{constraintsigma0b}).

Given these initial conditions, $\sigma $ can be expressed in terms of Hermite $\mathcal{H}$ and confluent hypergeometric (Kummer) functions  ${}_1F_1$ as
\begin{eqnarray}
\label{sigma combined long}
\nonumber\sigma(t) &=& \sigma_0
\left[\sqrt{3}\left(4+\sqrt{2}\sqrt{\rho_\phi/3} t_\text{eq}\right)
{}_1F_1\left(1-\frac{m_\sigma^2}{6\rho_\phi/3};\frac{3}{2};\frac{3}{8}(2\sqrt{2}+\sqrt{\rho_\phi/3} t_\text{eq})^2\right)\right.\\&&\nonumber\left.
\mathcal{H}_{\frac{m_\sigma^2}{3 \rho_\phi/3}}
\left(\sqrt{\frac{3}{2}}(t-t_\text{eq})\sqrt{\rho_\phi/3}-\sqrt{3}\right)\right.
-4\mathcal{H}_{\frac{m_\sigma^2}{3\rho_\phi/3}-1}
\left(-\frac{\sqrt{3}}{4}(4+\sqrt{2\rho_\phi/3} t_\text{eq})\right)\\\nonumber
&&\left.\left.
{}_1F_1\left(-\frac{m_\sigma^2}{6\rho_\phi/3};\frac{1}{2};\frac{3}{4}(\sqrt{2\rho_\phi/3}(t-t_\text{eq})-2)^2\right)\right]\right/\\ \nonumber&&
\left[\sqrt{3}\left(4+\sqrt{2\rho_\phi/3} t_\text{eq}\right)
{}_1F_1\left(1-\frac{m_\sigma^2}{6\rho_\phi/3};\frac{3}{2};\frac{3}{8}(2\sqrt{2}+\sqrt{\rho_\phi/3} t_\text{eq})^2\right)\right.\\\nonumber&&\left.
\mathcal{H}_{\frac{m_\sigma^2}{\rho_\phi/3}}
\left(-\frac{\sqrt{3}}{4}(4+\sqrt{2\rho_\phi/3} t_\text{eq})\right)
-4\mathcal{H}_{\frac{m_\sigma^2}{3 \rho_\phi/3}-1}
\left(-\frac{\sqrt{3}}{4}(4+\sqrt{2\rho_\phi/3} t_\text{eq})\right)
\right.\\&&\left.
{}_1F_1\left(-\frac{m_\sigma^2}{6\rho_\phi/3};\frac{1}{2};\frac{3}{8}(2\sqrt{2}+\sqrt{\rho_\phi/3} t_\text{eq})^2\right)
\right].
\end{eqnarray}
Since we are primarily interested in the leading order oscillation frequency and the amplitude's evolution, let us approach the problem from a more pragmatic perspective. The equation of motion (\ref{sigma eom}) is that of a damped simple harmonic oscillator with mass $m_\sigma$, justifying the ansatz
\begin{equation}
\label{sigma double}
\sigma\approx\sigma_0\left(\frac{t}{t_\text{i}}\right)^p \cos(m_\sigma(t-t_\text{i}))\,,
\end{equation}
\noindent
which solves (\ref{sigma eom}) to order $\mathcal{O}\left(m_\sigma/t_\text{eq}\right)$. Comparing to the full analytic solution, we find that the best fit for the exponent is $p\sim-1.5$. This exponent depends only weakly on $m_\sigma$, but changes somewhat with $t_\text{eq}$ (i.e.~it depends on $\sqrt{\rho_\phi/3}$ via $a_\text{eq}^{-2}=\rho_\phi/3$) and the size of the neighbourhood we are considering, $t_\text{f}-t_\text{i}$. Given the notation of Sec.~\ref{sec:history}, the initial and final time are
\begin{eqnarray}
t_\text{i}\equiv\sqrt{\frac{3}{\rho_\phi}}\,\text{arcsinh}(1/A)\quad,\quad t_\text{f}\equiv\sqrt{\frac{3}{\rho_\phi}}\text{arcsinh}(1/B)\,.\label{initialandfinaltimeinterm}
\end{eqnarray}
Since we chose $A=1/B=10$, we have
\begin{eqnarray}
t_\text{f}-t_\text{i}\approx 2.9 \, \sqrt{\frac{3}{\rho_\phi}} \approx 3.3\, t_\text{eq}\,.\label{def:deltat}
\end{eqnarray}

\begin{figure}[t]
        \begin{subfigure}[b]{0.5\textwidth}
                \centering
                \setlength\fboxsep{0pt}
                \setlength\fboxrule{0.5pt}
                \includegraphics[width=\textwidth]{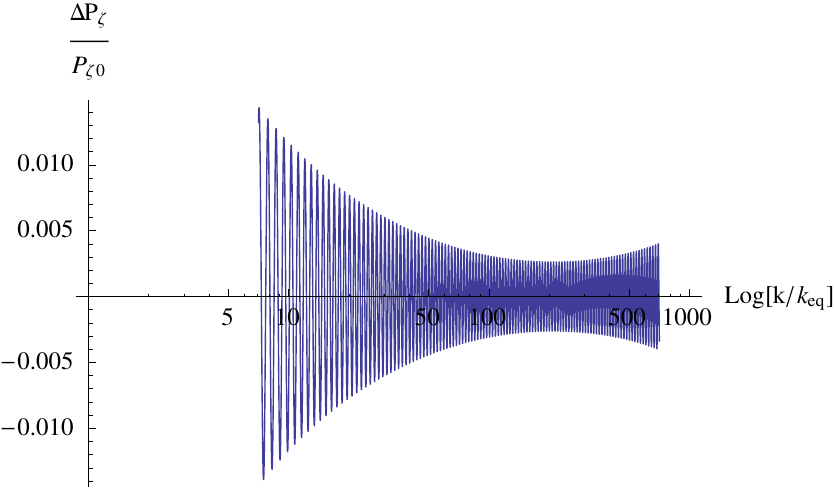}
                \caption{\small $m_\sigma=100\,\sqrt{\rho_\phi/3}$, $p=-1.54853$}
                \label{fig:power big time double}
        \end{subfigure}
        ~
        ~ 
        \begin{subfigure}[b]{0.5\textwidth}
                \centering
                \setlength\fboxsep{0pt}
                \setlength\fboxrule{0.5pt}        
                \includegraphics[width=\textwidth]{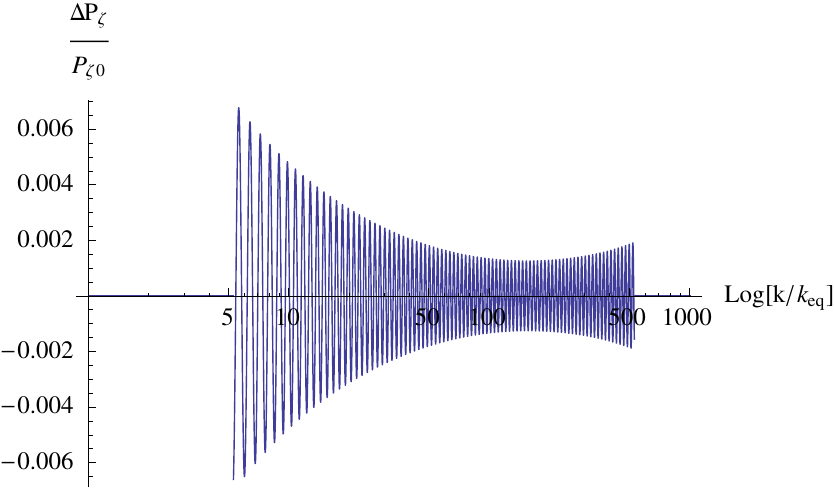}            
                \caption{\small $m_\sigma=75\, \sqrt{\rho_\phi/3}$, $p=-1.55151$}
                \label{fig:power eq time double}
        \end{subfigure}
        ~ 

\centering
        \begin{subfigure}[b]{0.5\textwidth}
                \centering
                \setlength\fboxsep{0pt}
                \setlength\fboxrule{0.5pt}
                \includegraphics[width=\textwidth]{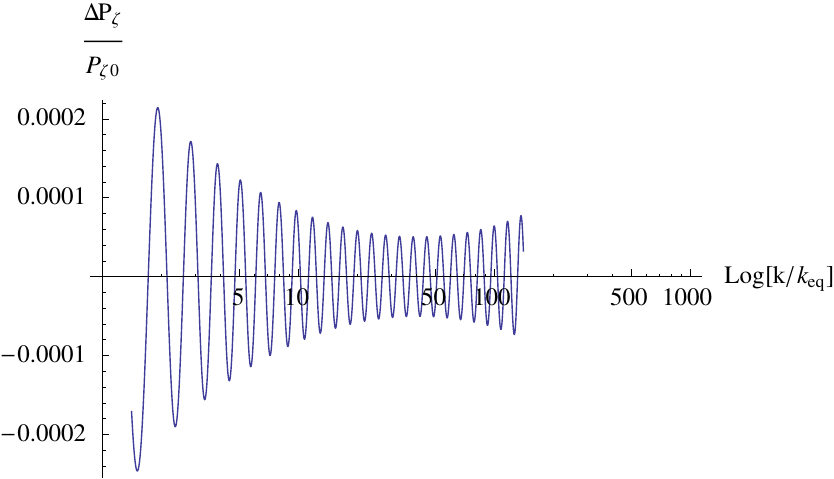}
                \caption{\small $m_\sigma=20\, \sqrt{\rho_\phi/3}$, $p=-1.53398$}
                \label{fig:power small time double}
        \end{subfigure}
        \caption{\small Power-spectrum correction in the intermediate regime for different masses of the $\sigma$-field, with $\sigma_0=0.01 $, and $A=10$, where $k_\text{eq}$ is the horizon scale at the time of equality. The power law index $p$ is determined by $m_\sigma$.}
        \label{fig:power double}
\end{figure}

From here on the computation proceeds as usual. The oscillatory component of the Hubble parameter is
\begin{equation}
\label{Hosci double}
H_\text{osci}=-\frac{\sigma_0^2}{12 }
\frac{m_\sigma}{H_\text{bgr}}
\,\frac{p}{t_\text{i}}\,\left(\frac{t}{t_\text{i}}\right)^{2 p-1}
\sin\left(2m_\sigma(t-t_\text{i})\right)\,,
\end{equation}
\noindent
and for a large enough ratio $m_\sigma/\sqrt{\rho_\phi/3}\gtrsim 15$, the dominant oscillatory component of the slow-roll parameters take the form 
\begin{eqnarray}
\label{eps osci double}
\epsilon_\text{osci}&=&
\frac{\sigma_0^2}{6 }\frac{m_\sigma^2}{H_\text{bgr}^3}
\frac{p}{t_\text{i}}
\left(\frac{t}{t_\text{i}}\right)^{2 p-1}
\cos\left(2 m_\sigma(t-t_\text{i})\right)\,,\\
\label{eta osci double}
\left(\epsilon \dot{\eta}\right)_\text{osci}&=&
-\frac{2 \sigma_0^2}{3 }\frac{ m_\sigma^4} {H_\text{bgr}^4}
\frac{p}{t_\text{i}} \left(\frac{t}{t_\text{i}}\right)^{2p-1}
\cos\left(2 m_\sigma(t-t_\text{i})\right)\,,
\end{eqnarray}
leading to the power-spectrum correction
\begin{eqnarray}
\label{power double}
\nonumber \frac{\Delta P_\zeta}{P_{\zeta}}\left(k\right)&=&
\frac{\sigma _0^2m_{\sigma }^2}{6  \rho_\phi/3}\frac{p}{t_\text{i}}
\int _{t_\text{min}}^{t_\text{max}} \left(\frac{t}{t_\text{i}}\right)^{2p-1}\frac{\sinh^2(\sqrt{\rho_\phi/3} t)}{\cosh (\sqrt{\rho_\phi/3} t)}
\cos \left(2 m_\sigma(t-t_\text{i})\right)\\&&
\left(\frac{\tanh^2(\tau)-2 k^2}{k}\sin(2k \tau )-2 \tanh(\tau) \cos (2k \tau) \right)d\,t
\end{eqnarray}
and shape function 
\begin{eqnarray}
\nonumber S\left(k_1,k_2,k_3\right)&=&
\mathcal{N}k_ 1 k_2 k_3\:
\frac{\sigma_ 0^2 m_\sigma^4\sqrt{\epsilon_\text{bgr}}p}
{24  H_{\text{bgr}}(t_0) t_i^m\sqrt{\rho_\phi/3}^3}
\int _{\tau_\text{min} }^{\tau_\text{max}}
\left(\frac{t}{t_i^m}\right)^{2p-1} \frac{\sinh ^3 (\sqrt{\rho_\phi/3} t)}{\cosh(\sqrt{\rho_\phi/3} t)}\\  &&
\left(2 K \cos (K \tau)+6 \,\text{sech} \left(\sqrt{\frac{\rho_\phi}{3}} t\right) \sin (K \tau)\right) \cos\left(2 m_\sigma(t-t_\text{i})\right)d\, t \,.
\end{eqnarray}
with $\mathcal{N}$ in (\ref{def:mathcalN}), which becomes
\begin{eqnarray}
\mathcal{N}=\left(\frac{2A^2}{A^2+1}\right)^{3/2}\approx 2^{3/2}\,,
\end{eqnarray}
where we used our approximate background solution in (\ref{scale intermediate}) and (\ref{Hbgrinterm}) as well as $A=10\gg 1$ from Sec.~\ref{sec:history} in the last step. 
The asymptotic behaviour of the power-spectrum is given by
\begin{eqnarray}
\label{power asym double}
\nonumber\frac{\Delta P_\zeta}{P_{\zeta}}(k)&\approx &
\frac{p\sqrt{\pi}}{12\,\text{arcsinh}(A^{-1})}
\sigma_0^2
\frac{\sqrt{\rho_\phi/3}}{m_\sigma}
k^\frac{3}{2}
\left(1+\frac{k^2\rho_\phi/3}{m_\sigma^2}\right)^{-\frac{7}{4}}
\left(\frac{\text{arcsinh}\left(\frac{k \sqrt{\rho_\phi/3}}{m_\sigma}\right)}{\text{arcsinh}(A^{-1})}\right)^{2p-1}\\
\nonumber &&
\left(
\left(2k^2\left(1+\frac{k^2 \rho_\phi/3}{m_\sigma^2}\right)-1\right)
\sin \left(
	2k\omega(k)
		-2 m_\sigma t_\text{i}-\frac{3\pi}{4}\right)\right.\\ 
&&\left.+
		2k\left(1+\frac{k^2 \rho_\phi/3}{m_\sigma^2}\right)^{\frac{1}{2}}
\cos \left(
	2k\omega(k)-2 m_\sigma t_\text{i}+\frac{\pi}{4}\right)\right),
\end{eqnarray}
with frequency
\begin{eqnarray}
\omega(k)\equiv \left(
		\frac{m_\sigma}{k\sqrt{\rho_\phi/3}}\text{arcsinh}\left(\frac{k\sqrt{\rho_\phi/3}}{m_\sigma}\right)
		+\,\text{arcsinh}\left(\frac{m_\sigma}{k\sqrt{\rho_\phi/3}}\right)\right)\,.
\end{eqnarray}
Similarly, the shape function becomes
\begin{eqnarray}
\label{shape double asym}
S(k_1, k_2, k_3)&\approx &
\mathcal{N}k_1 k_2 k_3
\frac{\sqrt{\pi}p}{16\,\text{arcsinh}(A^{-1})}
\sigma_0^2\left(\frac{K}{2}\right)^{\frac{7}{2}}
\left(1+\frac{K^2 \rho_\phi/3}{4 m_\sigma^2}\right)^{-\frac{5}{4}}
\\\nonumber&&
\left(\frac{\text{arcsinh}\left(\frac{K \sqrt{\rho_\phi/3}}{2 m_\sigma}\right)}{\text{arcsinh}(A^{-1})}\right)^{2p-1}\left(\cos \left(
	K\omega\left(K/2\right)-2 m_\sigma t_\text{i}-\frac{\pi}{4}\right)
		\right.\\\nonumber&&
\left.-\frac{K}{3}\left(1+\frac{K^2 \rho_\phi/3}{4 m_\sigma^2}\right)^{\frac{1}{2}}\sin \left(
			K\omega\left(K/2\right)-2 m_\sigma t_\text{i}-\frac{\pi}{4}\right)\right).
\end{eqnarray}

\begin{figure}[t]
        \begin{subfigure}[b]{0.5\textwidth}
                \centering
                \setlength\fboxsep{0pt}
                \setlength\fboxrule{0.5pt}
                \includegraphics[width=\textwidth]{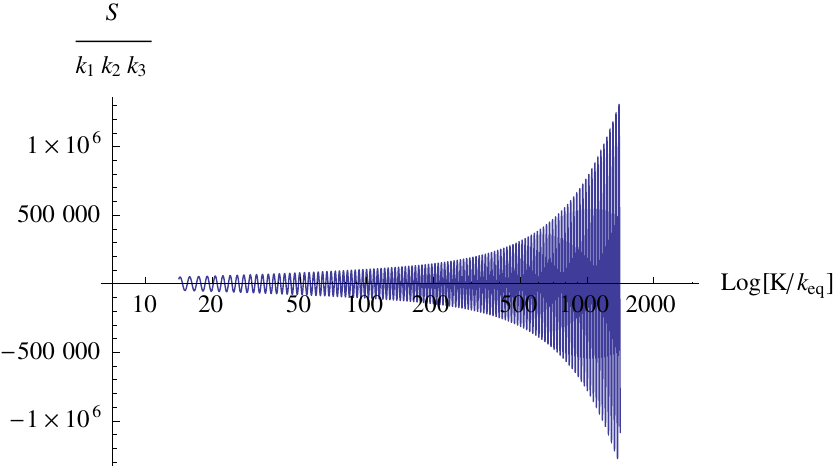}
                \caption{\small $m_\sigma=100\, \sqrt{\rho_\phi/3}$, $p=-1.54853$}
                \label{fig:bi small time double}
        \end{subfigure}
        ~
        ~ 
        \begin{subfigure}[b]{0.5\textwidth}
                \centering
                \setlength\fboxsep{0pt}
                \setlength\fboxrule{0.5pt}
                \includegraphics[width=\textwidth]{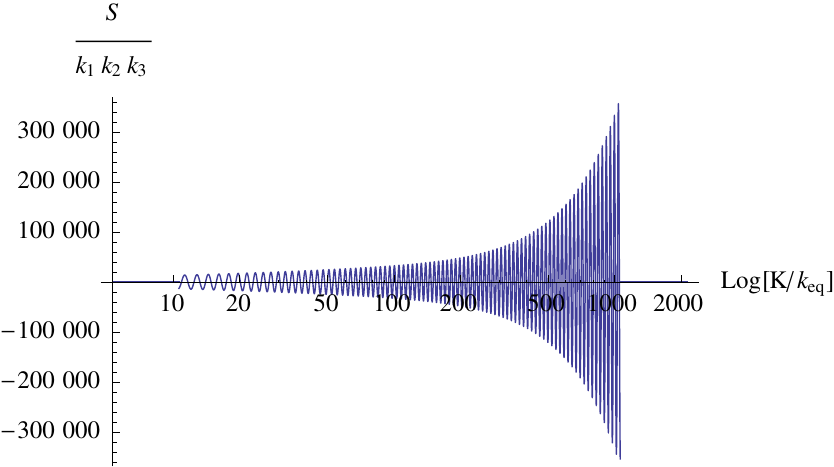}
                \caption{\small $m_\sigma=75\, \sqrt{\rho_\phi/3}$, $p=-1.55151$}
                \label{bi eq time double}
        \end{subfigure}
        ~ 

\centering
        \begin{subfigure}[b]{0.5\textwidth}
                \centering
                \setlength\fboxsep{0pt}
                \setlength\fboxrule{0.5pt}
                \includegraphics[width=\textwidth]{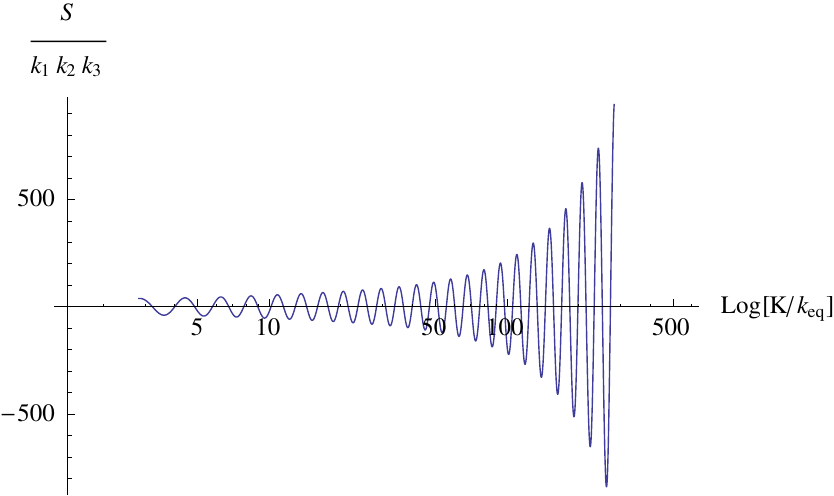}
                \caption{\small $m_\sigma=20\, \sqrt{\rho_\phi/3}$, $p=-1.53398$}
                \label{bi big time double}
        \end{subfigure}
        \caption{\small Shape function in the intermediate regime for different mass of the $\sigma$-field, with $\sigma_0=0.01 $, $t_0=t_\text{eq}$ so that $a(t_0)\,H_\text{bgr}(t_0)=\sqrt{2}$, where $k_\text{eq}$ is the horizon scale at the time of equality. The power law index $p$ is determined by $m_\sigma$.}\label{fig:bispec double}
\end{figure}
These signals, plotted in Fig.~\ref{fig:power double} and \ref{fig:bispec double} for a few exemplary parameters,  show a mixture of familiar features: firstly, the signal is bounded in the k-domain on both sides,
\begin{eqnarray}
\frac{m_\sigma}{A\sqrt{\rho_\phi/3}} \leq k\leq \frac{m_\sigma}{B\sqrt{\rho_\phi/3}}\,, \label{kdomainintermediate}
\end{eqnarray}
set by the duration of the intermediate regime (we chose $A=1/B=10$, and $k$ stands for $K/2$ in the shape function case). Note that signals from the curvature and inflationary regime are continuously connected in momentum space by the intermediate interval, with later times corresponding to higher wave numbers.
 
If $t_\text{f}-t_\text{i}<T=2\pi/m_\sigma$, no signal is produced in the power-spectrum (a resonance requires at least one full oscillation of $\sigma$), which translates into a lower bound on $m_\sigma$ given the duration in (\ref{def:deltat}).

Two unexpected features are evident in our analytic results:
\begin{enumerate}
\item The amplitude of the oscillations in the power-spectrum exhibits a local minimum and increases above a threshold wave number; we expect the amplitude to decrease monotonically, in order to provide a smooth transition to the inflationary regime, where the amplitude is monotonically decreasing, Fig.\ref{fig:power infl}.
\item The bi-spectrum does not show a local maximum, see Fig.~\ref{fig:bispec double}. As discussed in Sec.~\ref{sec:infl dom}, during the inflationary regime the bi-spectrum's envelope decreases monotonically, while it increases towards the end of the curvature-dominated regime, indicating a maximum in the intermediate one.
\end{enumerate}
Both of these anomalies are artefacts caused by using the approximation (\ref{sigma double}) for $\sigma(t)$, which underestimates the damping of oscillations for large $k$. In App.~\ref{sec: power-spectrum anomaly}, we revisit our computational strategy and explain these anomalies.

We conclude that oscillations in the power-spectrum are indeed monotonically decreasing through the three regime, while the oscillation amplitude of the bi-spectrum has a local maximum in the intermediate regime. Further, our analytic results are trustworthy at the lower end of the k-domain in (\ref{kdomainintermediate}).

\subsubsection{Another Case Study and Observational Prospects \label{sec:anothercasestudy}}
Using our toy model $V(\phi)=m_\phi^2\phi^2/2$ with $\phi=16$ and $m_\phi=\sqrt{8\pi}\times 10^{-6}$, let us estimate the maximally attainable amplitude of the power-spectrum correction induced by an initial excitation  $\sigma(t_{i})=\sigma_0$. Since the amplitude is monotonically decreasing, we compute the correction at the lower boundary of the k-domain,
\begin{eqnarray}
k_\text{min}=\frac{m_\sigma}{A\sqrt{\rho_\phi/3}}\approx \frac{m_\sigma}{m_\phi}\frac{\sqrt{6}}{A \phi}
\end{eqnarray}
with $A=10$. Further, the strongest constraint on $\sigma_0$ is (\ref{constraintsigma0a}),
\begin{eqnarray}
\sigma_0\ll \frac{m_\phi}{m_\sigma}\frac{1}{\delta}
\end{eqnarray}
with $\delta=0.1$. For example, if $m_\sigma=100\sqrt{\rho_\phi/3}$ we get $\sigma_0\ll 0.015$; in the plots of Fig.~\ref{fig:power double} and \ref{fig:bispec double}, $\sigma_0$ is close to this boundary.

Requiring that at least one $\sigma$-oscillation is completed in the intermediate regime, $2\pi/m_\sigma=T<t_\text{f}-t_\text{i}$, leads to
\begin{eqnarray}
\frac{m_\sigma}{m_\phi}&>&\frac{2\pi\phi}{\sqrt{6}}\frac{1}{\text{arcsinh}\left(\frac{1}{B}\right)-\text{arcsinh}\left(\frac{1}{A}\right)}\\
&\approx& 14.2\,,\label{lowerlimitmsigmaintermediate}
\end{eqnarray}
where we use the time interval in (\ref{initialandfinaltimeinterm}).

We can approximate the amplitude of the power-spectrum correction in (\ref{power asym double}) by evaluating it at $k=k_\text{min}$, since it decreases monotonically with $k$,
\begin{eqnarray}
\nonumber \frac{\Delta P_\zeta}{P_\zeta}&\sim& \sigma_0^2\frac{|p|\sqrt{\pi}}{6}\left(\frac{m_\sigma}{A\sqrt{\rho_\phi/2}}\right)^{\frac{5}{2}}
\approx  \sigma_0^2\frac{|p|\sqrt{\pi}}{6}\left(\frac{\sqrt{6}m_\sigma}{Am_\phi \phi}\right)^{\frac{5}{2}}\\
&\ll &\frac{|p|\sqrt{\pi}}{6\delta^2}\left(\frac{\sqrt{6}}{A\phi}\right)^{\frac{5}{2}}\sqrt{\frac{m_\sigma}{m_\phi}}
\approx 0.0013 \sqrt{\frac{m_\sigma}{m_\phi}}\,.
\end{eqnarray}
We focused on the sine term (it carries the larger amplitude), took $\text{arcsinh}(1/A)\approx 1/A$, ignored corrections of order $\mathcal{O}(1/A)$, used the upper bound on $\sigma_0$ in the second to last step and took $|p|\approx 3/2$ in the last one. Thus, a signal generated close to the lower limit in (\ref{lowerlimitmsigmaintermediate}), leads to a correction less then a quarter of a percent and for the largest possible mass, $m_\sigma\sim 1$, we  have 
\begin{eqnarray}
\frac{\Delta P_\zeta}{P_\zeta}\ll 0.57\,. \label{upperlimitPcaseB}
\end{eqnarray}
Hence, in the most optimistic case with the largest possible $\sigma_0$ allowed within our framework and a large mass, we might  observe an oscillation amplitude of a few percent. 

\section{Conclusion \label{sec:conclusion}}

Motivated by ubiquitous metastable vacua in the multi-dimensional landscape of string theory, we consider two-field open inflation as a model of the early universe, as proposed in \cite{Sugimura:2011tk}. Changes in the power-spectrum due to super-curvature fluctuations as well as a direct detection of curvature are known tell-tale signs of open inflation. In this article, we propose a third complementary probe: oscillations in correlation functions caused by oscillations in the tunnelling field $\sigma$ after tunnelling, but before slow-roll inflation. The oscillation frequency of corrections to the power-spectrum as well as the bi-spectrum is set by the inverse of the scale factor, enabling (in principle) a direct detection of the background evolution and thus the curvature dominated regime. To use resonances induced by oscillating heavy fields as direct probes of the background evolution was first put forth by Chen \cite{Chen:2011zf}, for instance, to discriminate inflation from proposed alternatives, 
such as bouncing cosmologies; since his results are not directly applicable to a curvature dominate universe, we revisit this computation in this article.

We follow Chen's approach to compute correlation functions in three regimes: curvature domination, inflation and an intermediate regime smoothly connecting the former two. After specifying the smooth background, we compute oscillatory corrections to the Hubble parameter and its derivatives induced by oscillations in $\sigma$. At the perturbed level, we ignore fluctuations in $\sigma$, paying special attention to the conditions under which such a truncation is feasible, leading to a strict upper bound on the allowed oscillation amplitude $\sigma_0$ in (\ref{constraintsigma0a}) and (\ref{constraintsigma0b}). This bound is not always imposed in Chen's work \cite{Chen:2011zf,Chen:2012ja}, but satisfied throughout this article, leading to corresponding limits on signals in correlation functions. For larger $\sigma_0$, which are physically possible, fluctuations in all fields need to be included, since they are strongly coupled to each other. 

With this caveat in mind, we compute correlation functions via the in-in formalism; the resulting integrals allow for resonances between mode-functions and oscillations in background quantities. 
Potentially resonating modes are deep inside the horizon ($k\gg aH$) where differences between the Bunch Davies state and the (excited) state after the tunnelling event are suppressed; hence, we can take the former one as an approximation. We evaluate the resulting integrals via the stationary phase approximation, leading to analytic expressions for the power-spectrum correction and the bi-spectrum. The relevant expressions are given in (\ref{power curv}), (\ref{power infl asym}), (\ref{power asym double}) and in (\ref{shape curv}), (\ref{shape infl asym}),   (\ref{shape double asym}) for signals generated during the curvature, inflationary and intermediate regime respectively. We recover Chen's results \cite{Chen:2011zf} in the inflationary regime.

We provide concrete case studies during curvature domination, Sec.~\ref{sec:casestudy}, and the intermediate regime, Sec.~\ref{sec:anothercasestudy}, to discuss the amplitude of signals and the feasibility  of  an observation. We find that signals caused by the initial excitation of $\sigma$ directly after the tunnelling event are too weak (within our computational strategy) to be detectable with current experiments. Larger signals caused by a bigger initial amplitude of 
$\sigma$ after the tunnelling event are conceivable, but require a full multi-field treatment at the perturbed level. However, a subsequent excitation can generate oscillations on top of the power-spectrum of a few percent, see (\ref{limitpowerecaseA}) and (\ref{upperlimitPcaseB}), without violating the applicability of our approximation scheme. To achieve such large signals, the initial amplitude of $\sigma$ as well as its mass need to be relatively large, and the signal needs to be generated shortly before or during the intermediate regime. Furthermore, slow-roll inflation should not last much longer than the minimal amount needed to bring down spatial curvature below observational limits. If several oscillations on top of the power-spectrum 
were detected, the distinct change in the frequency $\omega(k)$ could act as a fingerprint of open inflation. Regarding the detectability of oscillations on top of the CMB power-spectrum by WMAP and PLANCK we refer the interested reader to recent articles such as \cite{Meerburg:2011gd,Chen:2012ja,Saito:2012pd,Benetti:2012wu}.

In discussing the intermediate regime, we found a spurious local maximum in the power-spectrum, which turned out to be an artefact caused by our analytic approximations. A complementary direct numerical integration leads to a continuously decreasing amplitude of the oscillatory signal. On the other hand, a local maximum is present in the shape function, which  pinpoints the transition region from curvature domination to inflation. However, the amplitude of the bi-spectrum is suppressed. The latter may be increased in models that include a direct coupling between fields.
 
To conclude, resonances caused by oscillating, heavy  fields lead to oscillations on top of the power-spectrum and a distinct bi-spectrum, which can indeed provide a complementary, direct probe of two-field open inflation, if they were observed. A detection of open inflation would offer corroborating evidence for the string landscape and thus a window of opportunity to test one of the few generic predictions of string theory, see e.g.~\cite{Guth:2012ww}. Since the scientific gain of a detection is large, it appears prudent to us to intensify searches for such signals in the data,
in conjunction with searches for super-curvature fluctuations as well as spatial curvature, even if the conditions under which such signals are detectable with current experiments are not the most generic ones.

\acknowledgments
T.B.~is thankful for hospitality at the Laboratoire Astroparticule et Cosmologie (Paris), the Aspen Center for Physics (NSF Grant $\# 1066293$) and the Cern theory devision. We would also like to thank S.~Patil for crucial comments on the validity of truncating perturbations to the adiabatic sector,  X.Gao and S.~Mizuno for related discussions,  and D.~Battefeld for extensive comments on the draft. T.B.~would also like to thank A.~Eggemeier for exchanges on instantons and CdL tunneling.

\appendix

\section{Normalisation of the shape function \label{app:A}}

We briefly discuss the normalization of the shape function, since different choices can be found in the literature. Recall the definition of $S$ in (\ref{def shape function}),
\begin{equation}
\label{app: def shape function}
\left<\zeta^3\right>\equiv S(k_1, k_2, k_3) \frac{P_{\zeta}^2}{\prod\limits_{i} k_i^2} (2 \pi)^7 \delta^3 (\textbf{k}_1+\textbf{k}_2+\textbf{k}_3)\,,
\end{equation}
and note that the normalisation factor of $P_\zeta^2$  has its origin in the definition of the two-point correlation function $\left<\zeta^2\right>$ in (\ref{def power-spectrum}),
\begin{equation}
\label{app:def power-spectrum}
\left<\zeta^2\right>\equiv \frac{P_\zeta}{2 k_1^3} (2 \pi)^5 \delta(\textbf{k}_1+\textbf{k}_2)\,.
\end{equation}
Hence, the proportionality to $(k_2k_2k_3)^2$ of the shape function in (\ref{shape simple}),
\begin{equation}
\label{app:shape function}
S\left(k_1,k_2,k_3\right)=\frac{(k_1 k_2 k_3)^2}{(2 \pi)^4 P_{\zeta}^2} i\int _{\tau_\text{min}}^{\tau_\text{max}}d\tau  a^3\Delta(\epsilon  \dot{\eta})u_{k_1}^*u_{k_2}^*\frac{du_{k_3}^*}{d\tau }+\text{2 perm.} + c.c.\, ,
\end{equation}
is determined by the choice of normalisation and has no intrinsic meaning. For example, if we  chose to normalise $S$ with $P_\zeta^2(k)/(k_1k_2k_3)$ instead of $P^2_\zeta(k)/(k_1k_2k_3)^2$, we end up with a shape function, usually called $\mathcal{A}$, of the form
\begin{equation}
\label{app:shape function}
\mathcal{A}\left(k_1,k_2,k_3\right)=\frac{(k_1 k_2 k_3)}{(2 \pi)^4 P_{\zeta}^2} i\int _{\tau_\text{min}}^{\tau_\text{max}}d\tau  a^3\Delta(\epsilon  \dot{\eta})u_{k_1}^*u_{k_2}^*\frac{du_{k_3}^*}{d\tau }+\text{2 perm.} + c.c.\, ,
\end{equation}
 which simplifies for a Bunch-Davies mode function to
\begin{equation}
\label{app:shape simple}
\mathcal{A}\left(K\right)=\frac{1}{(a^3 H_\text{bgr}^4)|_{(t_0)}}
\frac{\sqrt{\epsilon_\text{bgr}}}{16}
 \int _{\tau_\text{min}}^{\tau_\text{max}}
 \frac{\Delta(\epsilon \dot{\eta})}{\epsilon_\text{bgr}^{3/2}}
\left(-2 K\cos(K\tau)+ 6 f(\tau)\sin(K \tau) \right) \,d\tau\,.
\end{equation}

\section{Anomalies in the Analytic Approximations\label{sec: power-spectrum anomaly}}

We wish to elaborate on the origin of the two unexpected features in our analytic results: the presence of a local minimum in the oscillation-amplitude of the power-spectrum correction and the absence of a local maximum in the oscillation-amplitude of the bi-spectrum in the intermediate regime.

\subsection{Power-Spectrum Correction} 
Since we performed an expansion around the time of equality and the threshold wave number is $k\gg k_\text{eq}$, it is conceivable that the observed increase is merely an artefact. 
 To test this hypothesis, we complement our analytic estimates by direct numerical integration.

\begin{figure}[t]
                \centering
                \includegraphics{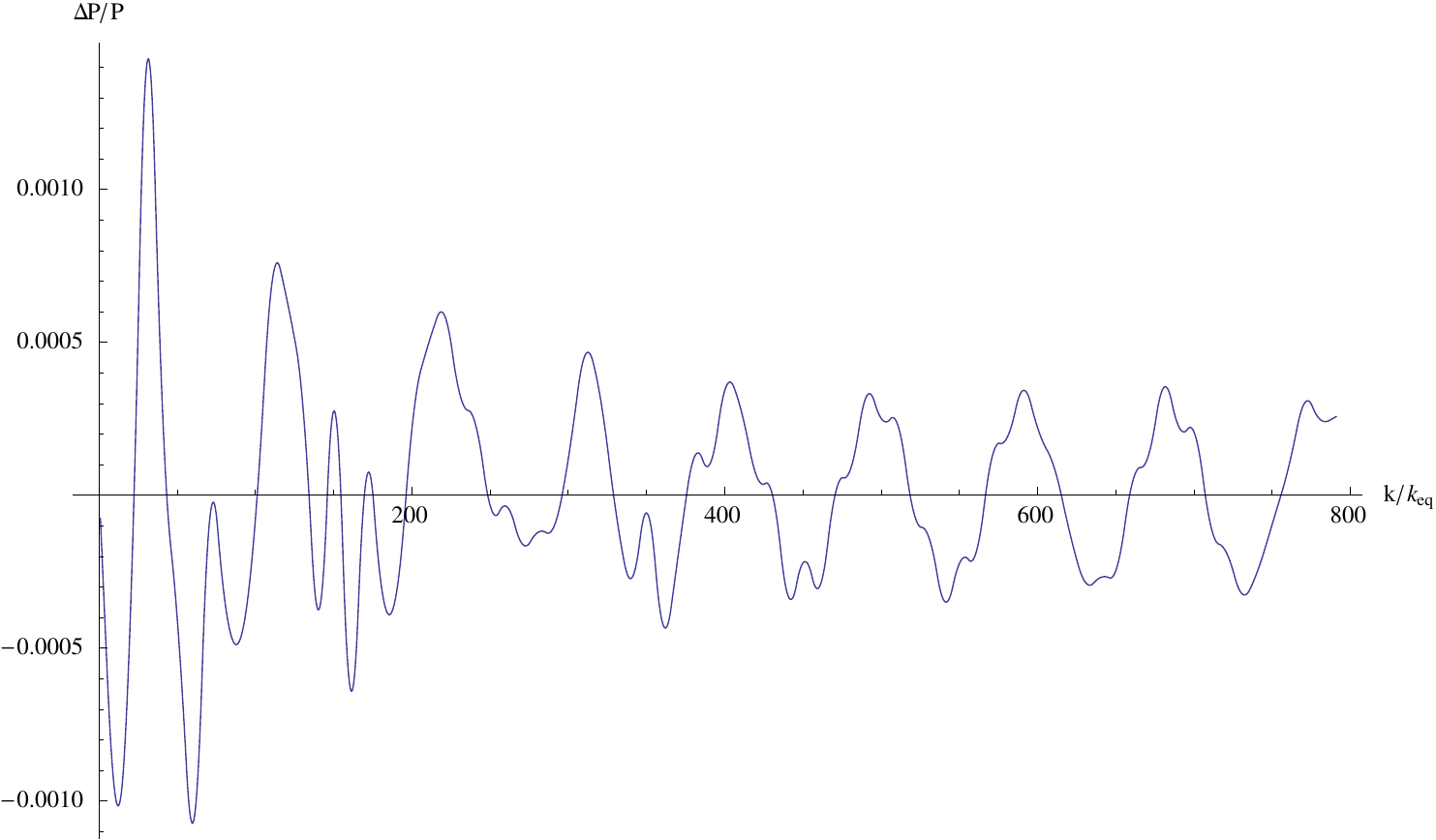}
                \caption{Numerical computation of the power-spectrum correction with $m_\sigma=75\, \sqrt{\rho_\phi/3}$ and  $\sigma_0=0.01 $. A monotonically decreasing amplitude of the oscillations is  visible (compare to Fig.~\ref{fig:power double}), leading to a smooth transition to the inflationary regime, Fig.~\ref{fig:power infl}. However, the numerical result is  contaminated by our imperfect subtraction of the smooth contribution, leading to irregularities in the oscillations.}
                \label{fig: double numerics}
        
\end{figure}

Firstly, we compute numerically the solution to the full equation of motion (\ref{sigma eom}) for $\sigma$ with $H_\text{bgr}=\sqrt{\rho_\phi/3}\coth (t\sqrt{\rho_\phi/3} )$. To separate $H_{\sigma \text{osci}}$ into smooth and oscillatory components, we take a moving average over a sufficient number of time steps, covering more than one oscillation. We obtain the slow-roll parameters, substitute them into (\ref{power simple}) and integrate numerically to compute the correction to the power-spectrum, see Fig.~\ref{fig: double numerics} for a representative example.

\begin{figure}[t]
                \centering
                \includegraphics{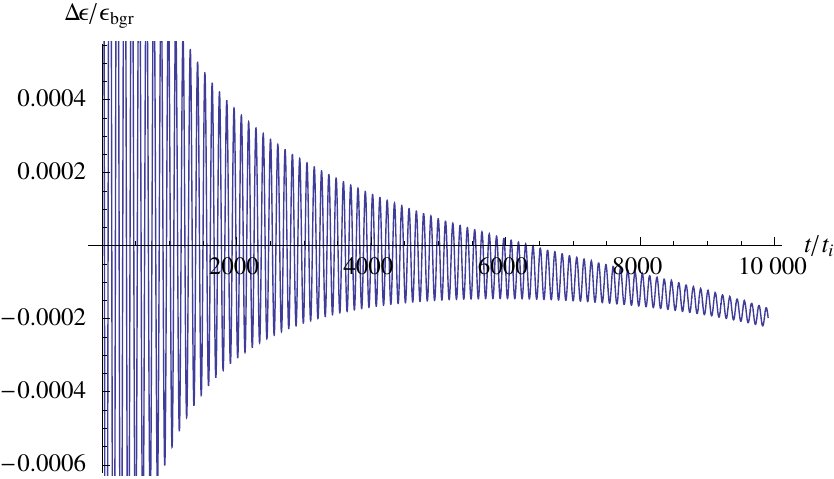}
                \caption{The oscillating component of the slow-roll parameter, computed numerically for $m_\sigma=100$ and $ \sqrt{\rho_\phi/3}$. The deviation from $\Delta \epsilon/\epsilon_\text{bgr}=0$ of the average is due to our imperfect subtraction of the leading order smooth component, which has no impact on our conclusions. }
                \label{fig:eps osci over eps}
\end{figure}

While the result is contaminated by our imperfect split into smooth and oscillatory components, we clearly observe a monotonically decreasing amplitude, confirming our suspicion that the minimum in the amplitude is indeed an artefact of the approximations used. To see exactly where the problem arises, consider the integral (\ref{power simple}). The key element is the contribution $\Delta \epsilon/\epsilon_\text{bgr}$, which simplifies to
\begin{equation}
\frac{\Delta \epsilon}{\epsilon_\text{bgr}}=-\frac{\dot{H}_\text{osci}}{\dot{H}_\text{bgr}}\,,
\end{equation}
where $\dot{H}_\text{osci}$ is the only unknown; according to the approximation in (\ref{sigma double}) as well as the numerical results, this function can be decomposed into a small oscillatory component with frequency $ \omega(t)$, overlaid onto a smoothly decaying envelope $A(t)$,  
\begin{eqnarray}
 \Delta \epsilon/\epsilon_\text{bgr}=A(t)/\dot{H}_\text{bgr} \sin(\omega(t))\,,
 \end{eqnarray} as shown in Fig.~\ref{fig:eps osci over eps} (without loss of generality we can choose $A(t)>0$).
Thus, the power-spectrum correction becomes\footnote{For simplicity, we neglect the contributions of the $f$-term in the power-spectrum integral, (\ref{power simple}).}
\begin{equation}
\frac{\Delta P}{P}(k)\sim \int\limits_{t_\text{min}}^{t_\text{max}}
\frac{A(t)k}{a \dot{H}_\text{bgr}}
\left( \sin(2 k \tau - \omega(t))+\sin (2 k \tau + \omega(t))\right) d t.
\end{equation}
Here, the extra factor of ``$a$'' in the denominator stems from the transformation to physical time. Due to the restrictions on the $\tau$-domain ($\tau<0$), only one of the sine terms contributes to the asymptotic behaviour of the integral. Since a minus sign can always be reabsorbed into the definition of $\omega(t)$ by exploiting the periodicity of the sine-function, we can freely take the contributing term to be the first one. Using the method of stationary phase, the envelope of the resulting oscillations in the power-spectrum is given by
\begin{equation}
\frac{\Delta P}{P}(k)_\text{env}\sim \frac{k_*A(t_*)}{a(t_*)\dot{H}_\text{bgr}(t_*)}
\chi(t_*)^{-1/2},
\end{equation}
with 
\begin{eqnarray}
\chi(t)\equiv\left.\left(\frac{\text{d}^2}{\text{d}t'^2}(2 k \tau - \omega(t'))\right)\right|_{t}
\end{eqnarray}
 and $t_*$  the solution of $k_*= a\, (d\omega/dt)/2\equiv g^{-1}(t_*)\equiv g^{-1}_*$. Substitution of the background variables yields
\begin{equation}
\label{power env}
\frac{\Delta P}{P}(k_*)_\text{env}\sim k_* A(g_*)(-a(g_*))\chi(g_*)^{-1/2}.
\end{equation}

Let's examine this expression carefully. Intuitively, the envelope should be a monotonically decreasing function of $k_*$, so as to smoothly connect with the beginning of the inflationary regime. Since monotonicity is preserved by composition but not by multiplication of functions, we note the origin and behaviour of the different multiplicative factors:
\begin{itemize}
\item $k_*$ comes from the mode function, i.e.~the Bunch-Davies vacuum oscillations; it trivially increases with $k_*$.
\item $-a(g_*)$ comes from the background interaction of the inflaton with the spatial curvature. It is monotonically increasing as a function of $t_*$, so that its overall behaviour is determined by $g_*$.
\item $A(g_*)$ - the amplitude of the $\sigma$ field oscillations is naturally decreasing with physical time due to the damping effect of the Hubble term. Its behaviour in momentum space is again determined by $g_*$.
\item $\chi(g_*)^{-1/2}$ - this term captures the strength of the resonance between the vacuum and $\sigma$-oscillations and its behaviour cannot be determined without a specific model.
\end{itemize}

\begin{figure}[t]
                \centering
                \includegraphics{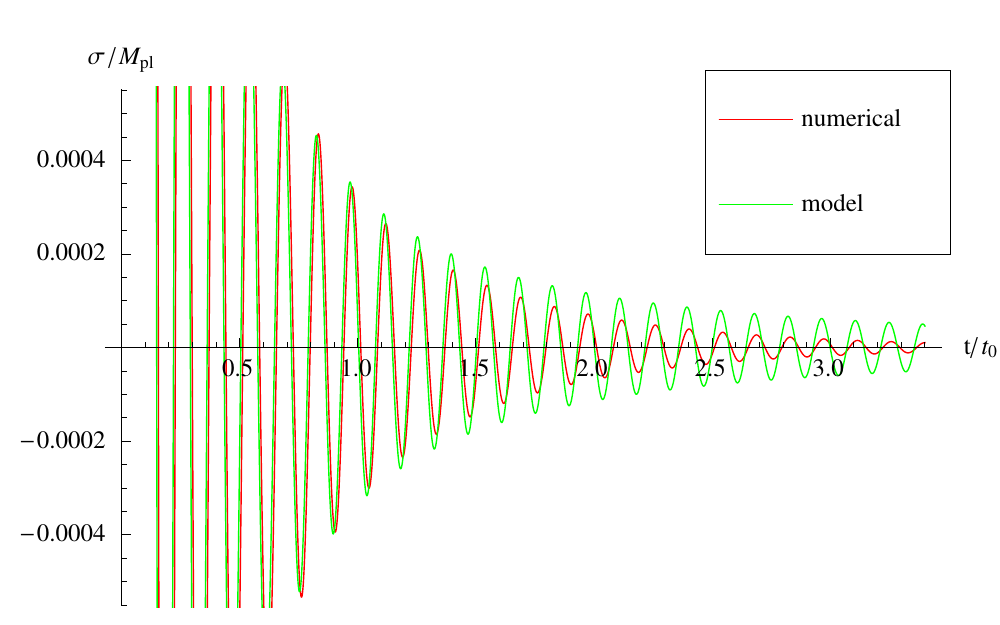}
                \caption{$m_\sigma=50\, \sqrt{\rho_\phi/3}$. The analytic approximation used in Sec.~\ref{sec:com contr} begins to deviate from the full numerical result at late times (high wave numbers) exemplified by under-damping and a phase shift in the $\sigma$-oscillations, leading to the unexpected increase in the power-spectrum signal, see Fig.~\ref{fig:power double}.}
                \label{fig:sigma ampl comp}
\end{figure}

To summarise, the monotonicity of the power-spectrum correction is determined primarily by the behaviour of the resonating wave mode $k_*=g_*^{-1}$ and the resonance mechanism $\chi_*$. In the case of the approximation (\ref{sigma double}), both $g_*$ and  $\chi(g_*)^{-1/2}$ increase monotonically. So, the only decreasing factor in (\ref{power env}) is the one related to the oscillation amplitude, $A(g_*)$. We conclude that the break of monotonicity observed in Fig.~\ref{fig:power double} has to be due to the insufficient damping of the $\sigma$-oscillations at high wave numbers (late times) in the power-law model. Indeed, (\ref{sigma combined long}) can be asymptotically expanded to reveal a factor of $\text{Exp} \left(\frac{3}{4}(\sqrt{2}+(t_i^m-t)\ \sqrt{\rho_\phi/3})^2 \right)$, which enhances the damping at late times. This is illustrated in Fig.~\ref{fig:sigma ampl comp}, which shows a comparison of the $\sigma$-oscillations obtained numerically and our analytic approximation.

We conclude that the power-spectrum correction does indeed monotonically decrease in the intermediate regime; only the approximation (\ref{power asym double}) breaks down beyond the stationary point of its envelope, leading to the spurious increase visible in Fig.~\ref{fig:power double}.

\subsection{Bi-spectrum}
 Applying the argument outlined above to the bi-spectrum dampens the monotonic increase of the amplitude at higher wave numbers as well, conceivably leading to the expected maximum. As the position and height of this maximum  depends sensitively on the chosen shape of the slow-roll potential, we are satisfied with our current understanding and leave a numerical computation to the future.

\end{document}